\documentclass[11pt, oneside]{article}   	
\usepackage{geometry}                		
\usepackage[parfill]{parskip}    		
\usepackage{graphicx}				
\usepackage[utf8]{inputenc}
\usepackage{amssymb,amsmath,amsthm}
\usepackage{algorithmic,algorithm}
\usepackage{hyperref}
\usepackage{natbib}
\usepackage{color}

\theoremstyle{remark}
\newtheorem*{remark}{Remark}

\DeclareMathOperator{\tr}{tr}

\title{Scalable Feature Matching Across Large Data Collections}

\author{David Degras} 

\begin{document}
\maketitle

\begin{abstract}
\noindent 
This paper is concerned with matching feature vectors in a one-to-one fashion across large collections of datasets.
Formulating this task as a multidimensional assignment problem with decomposable costs (MDADC),  
we develop extremely fast algorithms with  time complexity linear
in the number $n$ of datasets and space complexity a small fraction of the data size.  
These remarkable properties hinge on using the squared Euclidean distance as dissimilarity function,  
which can reduce ${n \choose 2}$ matching problems between pairs of datasets to $n$ problems  
and enable  calculating assignment costs on the fly.  
To our knowledge, no other method applicable to the MDADC possesses these linear scaling and low-storage properties 
necessary to large-scale applications.  
In numerical experiments, the novel algorithms outperform competing methods and show excellent computational and optimization performances.   
An application of feature matching to a large neuroimaging database is presented.  
The  algorithms of this paper are implemented in the R package \verb@matchFeat@ available at 
\href{https://github.com/ddegras/matchFeat}{github.com/ddegras/matchFeat}. 

\end{abstract}

\section{Introduction} 
\label{sec:introduction}

Matching objects across units (e.g. subjects, digital images, or networks) based on common descriptor variables is an ubiquitous  task 
in applied science. 
This problem, variously known as \emph{object matching}, \emph{feature matching},  \emph{data association}, or \emph{assignment problem}, 
is at the core of applications such as 
resource allocation  \citep{Pierskalla1968},
object tracking \citep{Thornbrue2010,Bar-Shalom2011,Dehghan2015,Rezatofighi2015,Smeulders2014,Wang2015},
object recognition \citep{Lowe2001,Belongie2002,Conte2004},
navigation systems \citep{Doherty2019},
image registration  \citep{LeMoigne2011,Ashburner2007},
optimization of communication networks \citep{Shalom2009}, 
connectomics in neuroscience \citep{Haxby2011,Vogelstein2015}, and more.

The impetus for this work is 
a task in functional neuroimaging which consists in matching collections of 
biomarkers (more precisely,  brain connectivity measures) between the subjects of a study. 
The matching process may serve in data exploration 
to provide new scientific insights and generate hypotheses.
It can also be a preliminary step in a group analysis 
to ensure meaningful comparisons across subjects. 
Key aspects of the matching problem under study are that: 
(i) the number of subjects and/or the number of biomarkers per subject may be large, 
posing computational challenges,  
(ii)  for two given subjects, each biomarker of one subject must be matched 
to at most one biomarker of the other 
(\emph{one-to-one matching}), 
and (iii) the matching must be consistent, i.e. transitive across subjects 
(for example, denoting subjects  by letters and biomarkers
 by numbers,  if  A1 is matched to B2 and  B2 to C3, then A1 must be matched to C3).  
This matching problem is not specific to neuroimaging and is applicable to the research fields mentioned above. 
It is generally relevant to \emph{multilevel} or \emph{hierarchical} analyses 
where outputs of a certain level of analysis must be matched before becoming inputs at the next level. 
This situation typically occurs when the outputs to be matched 
result from an unsupervised analysis such as clustering, segmentation, or dimension reduction. 

\paragraph{Problem formulation.} $^{}$
The matching problem at the core of this paper is as follows. 
Given $n$ set of vectors in $\mathbb{R}^p$ having the same size, say 
$\{ x_{11},\ldots , x_{1m}\}, \ldots, \{ x_{n1},\ldots , x_{nm}\}$,   
find permutations $\sigma_1,\ldots,  \sigma_n$ of the vector labels $\{1,\ldots,m\}$ 
 that minimize the sum of pairwise squared Euclidean distances within clusters $\{ x_{1\sigma_1(k)},\ldots , x_{n\sigma_n(k)} \}$ ($1\le  k \le m$).
Writing  $[r]=\{1,\ldots,r\}$ for a positive integer $r $ and 
letting $\mathbf{S}_m$ be the set of all permutations of $[m]$, 
the problem expresses as 
\begin{equation}\label{MDADC}
\min_{\sigma_1,\ldots,\sigma_n \in \mathbf{S}_m} \sum_{1\le i<j\le n} \sum_{k=1}^m 
\big\| x_{i\sigma_i(k)} - x_{j\sigma_j(k)} \big\|^2 
\end{equation}
where 
$\| \cdot\|$ denotes the Euclidean norm on $\mathbb{R}^p$. 
%


Problem \eqref{MDADC} is a sum-of-squares clustering problem 
with the constraint that each cluster must contain exactly 
one vector from each set $\{ x_{i1},\ldots , x_{im}\}$, $i \in [n]$. Identifying the $n$ sets with statistical units, 
this constraint guarantees that the obtained clusters reflect common patterns between units, 
not within units. For this reason, one-to-one feature matching is particularly suitable 
in applications where variations between units dominate variations within units. 


In problem \eqref{MDADC}, all statistical units  have the same number $m$ of vectors.  
It is natural to also set to $m$ the number 
of clusters to partition the vectors into. 
In practice however, statistical units may have unbalanced numbers of observations, 
say $m_1,\ldots,m_n$. It may also be desirable to group the observations in 
an arbitrary number of clusters, say $K$. 
Accordingly, a more general version of problem \eqref{MDADC} would be, 
for each $i\in [n]$, 
to assign vectors $ x_{i1},\ldots , x_{im_i}$ to $K$ clusters
 in a one-to-one fashion so as to minimize the total sum of pairwise squared Euclidean distances within clusters. 
 Here, one-to-one means that each unit $i$ can contribute at most one vector to any cluster:
 if $m_i=K$, each vector from unit $i$ is assigned to a cluster
 and each cluster contains exactly one vector from unit $i$;  
 if $m_i < K $, some clusters do not contain vectors from unit $i$;   
 and  if $m_i > K$, 
 some vectors from unit $i$ are not assigned to a cluster, i.e. they are unmatched. 
The matching problem  \eqref{MDADC} thus generalizes as 
\begin{equation}\label{objective unbalanced}
\min_{s_1,\ldots,s_n} \sum_{1\le i < j\le n} \sum_{k=1}^{K} 
\sum_{\substack{q \in [m_i],r \in [m_j] \\ s_i(q) = s_j(r) = k}}
 \left\| x_{i q} - x_{j r} \right\|^2 
\end{equation} 
where each $s_i$ is a map from the set  $[m_i]$ of vector labels to the set $\{0,\ldots,K\}$ of cluster labels 
where, by convention, labels of unassigned/unmatched vectors are mapped to the cluster label 0. 
The map $s_i$ is such that $s_i (q)=s_i(r)$ implies that $q=r$ or $s_i (q)=s_i(r)=0$. 
In other words the restriction of $s_i$ to $[m_i]\setminus s_i^{-1}(\{0\})$ must be an injective map. 
Problem \eqref{MDADC} is recovered when $m_1=\cdots = m_n = K :=m$, 
in which case $s_i = \sigma_i^{-1}$ for all $i\in [n]$.

For simplicity, only problem \eqref{MDADC} is treated in this paper. 
However,  the proposed matching methods extend to the general problem \eqref{objective unbalanced}. 
In complement to the model-free  problem \eqref{MDADC},  
a probabilistic approach to feature matching based on Gaussian mixtures is detailed in Section \ref{sec: GM}.

\paragraph{Related work.} 
Problem \eqref{MDADC} can be viewed through the prism of combinatorial optimization problems 
such as the \emph{minimum weight clique partitioning problem} in a complete $n$-partite graph, 
the \emph{quadratic assignment problem} \citep{Koopmans1957,Cela1998},
or the \emph{multidimensional assignment problem} (MAP) \citep[e.g.][]{Burkard2009}. 
The MAP formalism is well suited to this work and is recalled hereafter: 
\begin{equation}\label{MAP}
\min_{\sigma_1,\ldots,\sigma_n \in \mathbf{S}_m}
\sum_{k=1}^m c_{\sigma_1(k) \sigma_2(k) \cdots \sigma_n(k)}  
\end{equation}
where $(c_{a_1 a_2 \ldots a_n}) \in \mathbb{R}^{m^n}$ is an $n$-dimensional array containing 
the costs of assigning the feature vectors $x_{1 a_1}, \ldots , x_{n a_n}$ to the same cluster. 
Problem  \eqref{MDADC} is an instance of the MAP and, more precisely, it is a \emph{multidimensional assignment problem with decomposable costs} (MDADC) \citep[e.g.][]{Bandelt1994,Bandelt2004} 
because its assignment costs decompose as 
\begin{equation}\label{DC}
c_{a_1 a_2 \ldots i_n} =  \sum_{1\le i<j \le n}  d( x_{i a_i} , x_{j a_j} ) 
\end{equation}
where $d$ is a dissimilarity function.  
The squared Euclidean distance $d$ used in \eqref{MDADC} enables the development of  
highly efficient computational methods (see Section \ref{sec:methods}). 
The need for efficient computations comes from the exponential size $(m!)^n$ 
of the search domain $(\mathbf{S}_m)^n $
 and from the NP-hardness of  \eqref{MDADC} (when $n\ge 3$)  
 as a generalization of the 3D assignment problem of \cite{Spieksma1996}. 

The formidable literature on the MAP, 
which spans more than five decades and  multiple mathematical fields,
will not be reviewed here.
The interested reader may fruitfully consult 
\cite{Burkard2009} and \cite{Pardalos2013}.  
In fact, given the focus of the present work on computations,  
a broad review of the general MAP is not necessary. 
Indeed,  optimization methods for the MAP 
 \citep[e.g.][]{Karapetyan2011,Pierskalla1968,Poore1993,Robertson2001}
 are not computationally efficient for the special case of MDADC (and in particular \eqref{MDADC}),  
especially if the number $n$ of dimensions is large. 
We will  therefore only discuss the relevant MDADC literature.

\cite{Bandelt1994} provide simple  ``hub" and  ``recursive" heuristics for the MDADC 
\eqref{MAP}-\eqref{DC} along with their approximation ratios  
(worst-case bounds on the ratio of a method's attained objective to the optimal objective value).
The hub heuristic consists in selecting one dimension $i \in [n]$ 
of the MDADC as a ``hub" and matching all other dimensions to this one, 
i.e. finding for each dimension $j\ne i$ the assignment that minimizes the total cost with respect to $i$. 
The recursive heuristic starts by permuting the $n$ dimensions of the problem 
and then recursively finds the best assignment for the $i$th permuted dimension 
with respect to the $(i-1)$ first permuted dimensions ($i=2,\ldots, n$).   
\cite{Bandelt2004} enhance the heuristic methods of \cite{Bandelt1994} 
with local neighborhood search methods that attempt to improve a solution 
one or two dimensions at a time. They derive lower bounds for the minimum cost assignment 
based on a Lagrangian relaxation of the MDADC. 
\cite{Collins2012} also exploits the idea of improving solutions one dimension at a time 
in the general MAP \eqref{MAP} through a factorization technique. 
\cite{Kuroki2009} formulate \eqref{MDADC} as the problem of finding a clique cover 
of an $n$-partite graph with minimum edge weights. 
They express the clique cover problem with various  
 mathematical programs (integer linear, nonconvex quadratic, 
 integer quadratic, and second order cone) 
 which they tackle directly or after relaxation. 
They also provide approximation ratios and computational complexity bounds for their algorithms. 
\cite{Tauer2013} and \cite{Natu2020} solve Lagrangian relaxations of the integer linear program 
formulation of the MDADC, with an emphasis on efficient parallel computation 
in a Map-Reduce framework or with GPUs. They derive tight lower bounds 
to control the approximation error of their algorithms. 
%
 
As an alternative from the multidimensional assignment perspective,  
problem \eqref{MDADC} can be viewed as an instance of \emph{constrained clustering} 
or \emph{semi-supervised learning} \citep{Basu2009,Gancarski2020}. 
The constraint that  each unit $i\in[n]$ contributes exactly one feature vector to each cluster 
 can be rephrased as: 
two vector instances from the same unit cannot be assigned to the same cluster.  
This type of constraint, 
namely that certain pairs of instances cannot be assigned to the same cluster 
(``cannot link" constraint) or that certain pairs must be assigned to the same cluster 
(``must link" constraint), is called \emph{equivalence constraints} and 
can be handled by constrained $K$-means algorithms 
\citep{Wagstaff2001,Bilenko2004,Pelleg2007} 
or through constrained mixture models  \citep{Shental2003}.

Other tasks related to  problem \eqref{MDADC} but not directly relevant are object tracking, with applications in engineering and more recently in computer vision and artificial intelligence, and image registration, which plays a key role in image processing, object recognition, and remote sensing.   
The former involves a temporal dimension absent from \eqref{MDADC} 
whereas the latter involves many (and often noisy) features that are not matched one-to-one. 
%
Matching problems also have a long history in statistics and 
have been  a topic of intense scrutiny in machine learning in recent years
 \citep{DeGroot1976,Collier2016,Hsu2017,Pananjady2018}. 
 However, much of the research in these fields relevant to \eqref{MDADC} 
 deals with the case where $n=2$ and $m$ is large (asymptotically $m\to\infty$) 
 whereas we are chiefly interested in situations where $m$ is fixed 
 and $n$ is large ($n\to\infty$). 



\paragraph{Contributions.}

The methods  for the MDADC \eqref{MAP}-\eqref{DC}  discussed heretofore 
are applied in practice to problems of small size, say $n$ in the single digits or a few tens. 
Theoretical considerations as well as numerical experiments 
from this paper (see Sections \ref{sec:methods}-\ref{sec:simulations}) and from the literature indicate that these methods 
cannot handle large-scale problems with $n$ in the hundreds, thousands or more 
(at least, not in a reasonable time on a single computer). 
As a simple example, the ${n \choose 2} m^2$ costs  in \eqref{DC} are typically calculated and stored before 
starting the optimization, but even this preliminary step may exceed computer memory limits for large $n$ and/or $m$. 
In response to this methodological gap, our research aims to develop fast, scalable methods for matching feature vectors in a one-to-one fashion across a large number of statistical units. The main contributions of the paper are the following.

\begin{enumerate}
\item We develop very fast algorithms for solving the matching problem \eqref{MDADC}, 
that is, \eqref{MAP}-\eqref{DC}  with $d$ as the squared Euclidean distance. 
The three main algorithms (Sections \ref{sec:kmeans}-\ref{sec:bca}-\ref{sec:FW}) 
have iteration complexity $O(nm^3)$ and only take  a few iterations to converge, 
meaning that they scale linearly with $n$. 
In addition, they calculate assignment costs \eqref{DC} on the fly 
and have space requirements $O(mn+mp)$, a fraction of the data size $mnp$.  
We also present 
initialization methods and a refinement method  
(pairwise interchange). Further, we take a probabilistic view of \eqref{MDADC} as a constrained Gaussian mixture model  
and devise an efficient implementation of the Expectation-Maximization (EM)  algorithm. 

\item We provide a broad review of the diverse methods applicable to \eqref{MDADC}  (integer linear programming, various relaxations, constrained clustering) which rarely appear together in a paper. The novel algorithms are compared to these methods in numerical experiments and show excellent computation and optimization performances.

\item An R package \verb@matchFeat@ implementing all the algorithms of the paper is made 
 available at \href{https://github.com/ddegras/matchFeat}{github.com/ddegras/matchFeat}. 

\item  The matching problem \eqref{MDADC} is applied to a large database of neuroimaging data 
to study functional connectivity in the human brain. 
The data analysis confirms existing knowledge but also generates new insights, 
 thus demonstrating the practical usefulness of our approach.

\end{enumerate}

\paragraph{Organization of the paper.} 
Section \ref{sec:methods} introduces novel algorithms for the matching problem \eqref{MDADC}. 
In Section \ref{sec:simulations},  a numerical study assesses the novel algorithms and competing methods 
with respect to computation and optimization performance. 
Section \ref{sec:fmri} details an application of our matching approach to a large neuroimaging database (ABIDE) 
relating to autism spectrum disorders. 
 Concluding remarks are gathered in Section \ref{sec:discussion} 
 and additional details of the data analysis are provided in the Appendix.



\section{Novel algorithms for feature matching}
\label{sec:methods}

This section introduces novel algorithms for the matching problem \eqref{MDADC}. 
The first four are local search methods that aim to improve existing solutions.  
At the end of the section, we discuss initialization techniques 
for the local search methods.

\subsection{$K$-means matching} 
\label{sec:kmeans}

For a given $n$-uple of permutations $\sigma = (\sigma_1,\ldots,\sigma_n ) \in (\mathbf{S}_m)^n$, let $\overline{X}_{\sigma} $ be the average matrix 
of the permuted data with columns  $\overline{x}_{\sigma ,k } = \frac{1}{n} \sum_{i=1}^n x_{i \sigma_i(k)} $ 
for $1\le k \le m $.  Problem \eqref{MDADC} is equivalent to 
\begin{equation}\label{objective 2}
\min_{\sigma_1,\ldots,\sigma_n} \sum_{i=1}^{ n} \sum_{k=1}^m \left\| x_{i\sigma_i(k)} - \overline{x}_{\sigma,k} \right\|^2
\end{equation}

The following method  adapts the standard $K$-means clustering algorithm \citep{Lloyd1982} to the matching problem \eqref{objective 2}.  
\begin{enumerate}
\item Initialize $\sigma = (\sigma_1,\ldots,\sigma_n )$ to some arbitrary value, for example 
$\sigma = (\mathrm{Id}_{[m]}, \ldots, (\mathrm{Id}_{[m]} )$. 
Calculate the average matrix $\overline{X}_{\sigma} $ 
and the objective value  \eqref{objective 2}.
\item Given the average matrix $\overline{X}_{\sigma} $: 
for $ 1\le i\le n,$ find the permutation $\sigma_i$ that minimizes 
$ \sum_{k=1}^m \| x_{i\sigma_i(k)} - \overline{x}_{\sigma,  k} \|^2$. 
Update the solution to $\sigma = (\sigma_1,\ldots,\sigma_n )$. 
\item Given $\sigma $: calculate the average matrix $\overline{X}_{\sigma} $
and the objective value \eqref{objective 2}. If the objective has not decreased 
from the previous iteration, terminate the execution and return $\sigma$. Else go back to step 2. 
\end{enumerate}

Steps 2 and 3 above are non-increasing in the objective \eqref{objective 2}. 
For this reason, and due to the finiteness of the search space, the proposed approach converges in a finite number of iterations. Like the $K$-means, it only finds a local minimum of \eqref{objective 2}. 

Concerning computations, step 3  can be performed in $O(nm)$ flops.  
Step 2, which consists of $n$ separate optimizations, is the computational bottleneck. 
Observe that 
\begin{align*}
 \sum_{k=1}^m \| x_{i\sigma_i(k)} - \overline{x}_{\sigma,  k} \|^2 
 & =  \sum_{k=1}^m \| x_{i\sigma_i(k)} \|^2 - 2 \sum_{k=1}^m  \langle x_{i\sigma_i(k)} , \overline{x}_{\sigma,  k}\rangle
 +  \sum_{k=1}^m \|  \overline{x}_{\sigma,  k} \|^2 \\
 & =  \sum_{k=1}^m \| x_{ik} \|^2 - 2 \sum_{k=1}^m  \langle x_{i\sigma_i(k)}, \overline{x}_{\sigma,  k}\rangle
 +  \sum_{k=1}^m \|  \overline{x}_{\sigma,  k} \|^2 
 \end{align*}
where $\langle \cdot , \cdot \rangle$ denotes the Euclidean scalar product. That is,  
the minimization of $ \sum_{k=1}^m \| x_{i\sigma_i(k)} - \overline{x}_{\sigma,  k} \|^2$ 
(with respect to $\sigma_i \in \mathbf{S}_m$) 
is equivalent to 
\begin{equation}\label{LAP k-means}
\max_{\sigma_i \in \mathbf{S}_m} \sum_{k=1}^m \langle x_{i\sigma_i(k)} , \overline{x}_{\sigma,  k} \rangle
\end{equation}
Problem \eqref{LAP k-means} is an instance of the well-known \emph{linear assignment problem} (LAP)
\citep[e.g.][Chap. 4]{Burkard2009}.
 After calculating the assignment matrix 
 $A = (\langle \overline{x}_{\sigma,  k} ,  x_{il}\rangle )_{1\le k,l \le m}$,  the LAP  \eqref{LAP k-means}
 can be solved for example with the Hungarian algorithm \citep{Kuhn1955,Munkres1957}. 
Efficient implementations of the Hungarian algorithm have complexity $O(m^3)$. 
 
The $K$-means matching algorithm is summarized hereafter. 
The objective value in \eqref{objective 2} is denoted by $F(\sigma)$. 
\begin{algorithm}
\caption{$K$-Means Matching}
\label{alg:kmeanslike} 
\begin{algorithmic}[1]
\REQUIRE   $X_1,\ldots, X_n \in \mathbb{R}^{p\times m}$,   
$\sigma=(\sigma_1,\ldots, \sigma_{n}) \in (\mathbf{S}_m )^n$
\STATE  $\overline{x}_{\sigma ,k } \leftarrow (1/n) \sum_{i=1}^n x_{i \sigma_i(k)}  \, (1\le k \le m)$, 
 $F_{new} \leftarrow F(\sigma)$
\REPEAT
\STATE $F_{old} \leftarrow F_{new}$
\FOR{$i=1, \ldots, n$}
\STATE Solve the LAP \eqref{LAP k-means} 
and  call $\sigma_i^+$ a solution. 
\STATE $\sigma_i \leftarrow \sigma_i^+$
\ENDFOR
\STATE $\sigma \leftarrow (\sigma_1,\ldots, \sigma_{n})$
\STATE   $\overline{x}_{\sigma ,k } \leftarrow (1/n) \sum_{i=1}^n x_{i \sigma_i(k)}  \, (1\le k \le m)$, 
 $F_{new} \leftarrow F(\sigma)$
\UNTIL{$F_{new} \ge F_{old}$}
\end{algorithmic}
\end{algorithm}

\bigskip
\begin{remark} 
If $p=1$, the matrices $X_i$ are row vectors and the $x_{ik}$ are scalars. In this case, step 2 of the proposed method is extremely simple. Indeed for each $1\le  i \le n$, the sum $ \sum_{k=1}^m   x_{i\sigma_i(k)}  \overline{x}_{\sigma  ,k} $ 
 is maximized when the $x_{ik}$ and $ \overline{x}_{\sigma,k}$ are matched by rank.  
More precisely, take  any $s_i \in \mathbf{S}_m$ such that $x_{i s_i (1)} \le \cdots \le x_{i s_i(m)}$ and any $s \in  \mathbf{S}_m$ such that $ \overline{x}_{\sigma, s(1)}\le \ldots \le  \overline{x}_{\sigma, s(m)} $. Then $\sigma_i = s_i \circ s^{-1}$ maximizes the sum. In other words, the optimal permutations $\sigma_i$  are simply obtained by sorting the components of the $X_i$ and  $\overline{x}_{\sigma}$ (computational complexity $O(nm \log m)$).
\end{remark}

\subsection{Block coordinate ascent method}
\label{sec:bca}

For convenience problem \eqref{MDADC} is rewritten here using permutation matrices $P_1,\ldots,P_n$ instead of permutation functions $\sigma_1,\ldots,\sigma_n$.  Each  $P_i$ ($1\le i \le n$) is a square matrix with entries in $\{0,1\}$ such that each row and each column contains the value 1 exactly once.  
Let $\Pi_{m}$ be the set of all $m\times m$ permutation matrices. Problem \eqref{MDADC} expresses as the binary quadratic assignment problem
\begin{equation}\label{objective P}
\min_{P_1,\ldots,P_n \in \Pi_{m}} \sum_{ i=1}^{ n} \sum_{ j=1}^{ n} \left\| X_iP_i - X_j P_j  \right\|_{F}^2
\end{equation}
where 
$\| \cdot\|_F$ denotes the Frobenius norm ($\| X\|_F = \langle X , X \rangle_F^{1/2} =  
(\mathrm{tr}(X'X))^{1/2}$ with $\tr (\cdot)$ the trace operator). 
By expanding the squared Frobenius norm in the objective and noting that column permutations do not change the Frobenius norm of  matrix, we have
\begin{align*}
 \sum_{ i=1}^{ n} \sum_{ j=1}^{ n} \left\| X_iP_i - X_j P_j  \right\|_{F}^2 
& =  \sum_{ i=1}^{ n} \sum_{ j=1}^{ n} \left( \| X_iP_i\|_F^2 + \|X_j P_j \|_F^2 - 2 \langle X_iP_i , X_jP_j \rangle_F \right) \\
& =  \sum_{ i=1}^{ n} \sum_{ j=1}^{ n} \left( \| X_i \|_F^2 + \|X_j  \|_F^2 \right)  - 2  \bigg\| \sum_{i=1}^n X_iP_i \bigg\|_F^2  .
\end{align*}
Discarding  terms that do not depend on $P_1,\ldots,P_n$, 
problem   
\eqref{objective P} is equivalent to 
 \begin{equation}\label{objective P max}
\max_{P_1,\ldots,P_n \in \Pi_{m}} \bigg\| \sum_{i=1}^n X_iP_i \bigg\|_F^2 .
\end{equation}
The maximization problem \eqref{objective P max} can be handled one matrix $P_i$ at a time ($1\le i\le n$), that is, by \emph{block coordinate ascent} \citep[BCA, e.g.][]{Wright2015}. 
Given a current solution $(\hat{P}_1,\ldots, \hat{P}_n)$ and an index $i$, 
all matrices $\hat{P}_j, \, j\ne i$ are fixed and the task at hand is  
$$ \max_{P_i \in \Pi_{m} } \bigg\| X_i P_i + \sum_{\substack{  1 \le j \le n \\ j\ne i }} X_j \hat{P}_j \bigg\|_F^2 $$
which, after expansion, is equivalent to the linear assignment problem
\begin{equation}\label{LAP}
\max_{P_i \in \Pi_{m} } \Big\langle P_i , X_i' \sum_{ j\ne i } X_j \hat{P}_j \Big\rangle_F .
\end{equation}
As  mentioned in Section \ref{sec:kmeans},  \eqref{LAP} can be efficiently solved with the Hungarian algorithm. 
The permutation matrix $\hat{P}_i$ is then updated to a solution of \eqref{LAP}. 
This operation is repeated with the index $i$  sweeping through the set $[n]$ 
until no further increase in the objective \eqref{objective P max} has been achieved in a full sweep. 
 Given that each update of a $\hat{P}_i$ is non-decreasing in the objective \eqref{objective P max} and that  
the search domain $\Pi_{m}^n$ is finite, the algorithm is guaranteed to converge in a finite number of steps. 
Popular methods for sweeping through $[n]$ 
include the cyclical order (also known as the Gauss-Seidel rule), 
random sampling, random permutation of $[n]$, and greedy selection.

 The BCA algorithm is summarized hereafter. 
The objective function in \eqref{objective P max} is denoted by $F$. 
For simplicity the sweeping order is taken to be cyclical 
but any other sweeping method can be used. 

\begin{algorithm}
\caption{Block Coordinate Ascent}
\label{alg:bca} 
\begin{algorithmic}[1]
\REQUIRE   $X_1,\ldots, X_n \in \mathbb{R}^{p\times m}$,   
$P_1,\ldots, P_{n} \in \Pi_{m} $.
\STATE $S \leftarrow   \sum_{i=1}^{n} X_i P_i $, $F_{new} \leftarrow \| S \|_F^2$
\REPEAT
\STATE $F_{old} \leftarrow F_{new} $
\FOR{$i=1,\ldots, n$}
\STATE $S_i \leftarrow S - X_i P_i $
\STATE Solve the LAP 
$\max_{P_i \in \Pi_{m} } \big\langle P_i , X_i' S_i \big\rangle_F $ 
 and call $P_i^+$ a solution. 
\STATE $P_i \leftarrow P_i^+$, 
 $S \leftarrow S_i + X_i P_i$
\ENDFOR
\STATE $F_{new} \leftarrow \| S \|_F^2$
\UNTIL{$F_{new} \le F_{old}$}
\end{algorithmic}
\end{algorithm}

Algorithm \ref{alg:bca} can be viewed as a special case of the local search algorithm LS1 of \cite{Bandelt2004}. 
 The LS1 algorithm is more general in that   
it uses an arbitrary dissimilarity function $d$ in the MDADC \eqref{MAP}-\eqref{DC}. 
The computational price to pay for this generality is that  for each block update ($i\in [n]$) 
  the assignment matrix $A_i = ( \sum_{j \in [n]\setminus \{i\}} d(x_{j\sigma_j(k)} , x_{il} ))_{1\le k,l \le m}$ 
must be calculated from scratch  in $O(nm^2)$ flops. 
Hence the LS1 method has iteration complexity $O(n^2m^2)$ 
(one iteration meaning one sweep through $[n]$) which may be prohibitive for large $n$.  
In comparison, the squared Euclidean distance $d = \| \cdot \|^2$ employed in the BCA method
enables efficient computation of $A_i $ in $O(m^2)$ complexity 
by keeping track of the running sum $\sum_{i=1}^{n} X_iP_i$ with rank-1 updates. 
Accordingly, the BCA method has iteration complexity $O(nm^3)$ linear in $n$.   
A variant  of the BCA method using asynchronous parallel 
updates of the matrices $\hat{P}_i$ (the so-called Jacobi update) can further reduce 
the iteration complexity, although convergence properties of this approach are not clear.

\subsection{Convex relaxation and Frank-Wolfe algorithm} 
\label{sec:FW}

In the previous section, problem \eqref{objective P max} was solved  one permutation matrix $P_i$ at a time 
while keeping the other $P_j$ ($j \ne i $) fixed. 
As an alternative, one may relax this problem to the set $\mathcal{D}_m$ of doubly stochastic matrices of dimensions  $m\times m$, which is the convex hull of $\Pi_m$. (As a reminder, a doubly stochastic matrix is a square matrix with elements in 
 $[0,1]$  whose rows and columns all sum to 1.) The relaxed problem is 
 \begin{equation}\label{cvx relax}
\max_{P_1,\ldots,P_n \in \mathcal{D}_m } \Big\| \sum_{i=1}^n X_iP_i \Big\|_F^2 .
\end{equation}
Although this relaxation leads to an indefinite program (i.e. maximizing a convex quadratic form), it is the correct way to relax \eqref{objective P}-\eqref{objective P max}. In contrast, directly relaxing \eqref{objective P} (to $\mathcal{D}$) would produce a convex program that is computationally simpler but does not provide tight bounds \citep{Lyzinski2016}. 

The Frank-Wolfe algorithm \citep{Frank1956} is an excellent candidate for this maximization. 
Indeed the gradient of \eqref{cvx relax} is straightforward to compute. Denoting by $F$ 
the objective function of \eqref{cvx relax}, 
the partial derivatives are simply $ \partial F / \partial P_i = X_i' \sum_{j=1}^n X_j P_j$ ($1\le i \le n$). 
In addition, the associated linear program 
\begin{equation}
\max_{Q_1,\ldots,Q_n \in \mathcal{D}_m }  \sum_{i=1}^n \Big\langle Q_i , X_i' \sum_{j=1}^n X_j P_j \Big\rangle_F
\end{equation}
which provides the search direction $(Q_1,\ldots,Q_n)$ for the next algorithm iterate 
is easily solvable as $n$ separate linear assignment problems (LAP).  
Although each LAP is solved over $ \mathcal{D}_m$, 
 Birkhoff-von Neumann's theorem guarantees that a solution can be found in $\Pi_m$, 
 a property referred to as the integrality of assignment polytopes \citep{Birkhoff1946,vonNeumann1953}.  

Having found the search direction, it remains to select the step size $\alpha \in [0,1]$. 
This is often done with a line search: $\max_{\alpha \in [0,1]} F(P + \alpha (Q-P))$ 
where $P=(P_1,\ldots, P_n )$ and $Q = (Q_1, \ldots , Q_n)$. 
The expression to maximize is  a quadratic polynomial in $\alpha$ 
with leading coefficient $  \| \sum_{i=1}^n X_i(Q_i -P_i)  \|_F^2 \ge 0 $. 
Accordingly, the maximum over $[0,1]$ is attained either at $\alpha =1$  
or at $\alpha = 0$. In the former case, the algorithm takes a full step in the direction $Q$ 
whereas in the latter case, the current solution cannot be improved upon and the algorithm ends. 
Interestingly, the iterates generated by the Frank-Wolfe algorithm for 
problem \eqref{cvx relax} stay in $\Pi_m$ although in principle, 
they could also explore the interior of $\mathcal{D}_m$. 
This is a consequence of the integrality of the search direction $Q$  
and of the  line search method for a 
quadratic objective,  
which make the step size $\alpha$ equal to 0 or 1.

\begin{algorithm}
 \caption{Frank-Wolfe} 
 \label{alg:Frank-Wolfe}
 \begin{algorithmic}[1]
 \REQUIRE{$X_1,\ldots, X_m \in \mathbb{R}^{p\times m}$, $P_1,\ldots,P_n\in \mathcal{D}_m$}
 \STATE $S \leftarrow   \sum_{i=1}^{n} X_i P_i $, $F_{new} \leftarrow \| S \|_F^2$
 \REPEAT
\STATE $S' \leftarrow 0, \ F_{old} \leftarrow F_{new}$ 
\FOR{$i =1$ to $n$}
\STATE Solve the LAP $\max_{Q_i \in\mathcal{D}_m}  \big\langle Q_i , X_i' S \big\rangle_F $   
 and call $Q_i $  a solution.  
\STATE $ S' \leftarrow S' + X_i Q_i $
\ENDFOR 
\STATE $F_{new} \leftarrow  \|S'   \|_F^2 $ 
\IF{ $F_{new} > F_{old} $}
\STATE $P_i \leftarrow Q_i \ (1\le i \le n), \ S \leftarrow S' $
\ENDIF
\UNTIL{$F_{new} \le F_{old}$}
 \end{algorithmic}
\end{algorithm}
 
%

%
%
%

 \subsection{Pairwise interchange heuristic}
\label{sec:2exchange}

The BCA algorithm of Section \ref{sec:bca} attempts to improve an existing solution to \eqref{MDADC} one permutation $\sigma_i$ at a time. In other words, at each iteration, it changes all assignments $\sigma^l = (\sigma_1(l), \ldots, \sigma_n(l))$ ($ 1 \le l \le m$) 
in a single dimension.   \cite{Karapetyan2011} call this approach  a \emph{dimensionwise heuristic}.  
Another strategy called the \emph{interchange} or \emph{$k$-exchange heuristic}
is to change a few assignments  (typically, $k=2$ or $k=3$)  in all dimensions by element swaps 
\citep[e.g.][]{Balas1991,Robertson2001,Oliveira2004}. 
Here we consider the 2-assignment exchange algorithm (Algorithm 3.4) 
of \cite{Robertson2001} for the general MAP \eqref{MAP} 
and adapt it to problem \eqref{MDADC}. 
In this algorithm,  
given two assignments, 
the search for the best interchange 
is done exhaustively. 
This involves accessing as many as $2^n - 1$ candidate assignments for element swaps 
and comparing their costs, which is reasonable in the general MAP provided that: 
(i) costs are precalculated, (ii) $n$ is small, and 
(iii) candidate assignments for exchange are easily found among all feasible assignments.   
However, for moderate to large $n$, 
and in the context of problem \eqref{MDADC}
where assignment costs are not precalculated, 
the calculation and exhaustive search 
of $2^{n}-1$ interchange assignment costs 
for at least each of ${m \choose 2}$ 
candidate pairs of assignments are untractable. 
 We will show that in problem \eqref{MDADC}, 
the pairwise interchange heuristic 
can be efficiently solved as a 
binary quadratic program.

Given a solution $\sigma = (\sigma_1,\ldots, \sigma_n)$ to \eqref{MDADC} 
and two associated assignments $\sigma^q$ and $\sigma^r$
($1 \le q < r \le m$), 
the basic problem of pairwise interchange 
is to improve the objective in  \eqref{MDADC} 
by interchanging elements between these assignments, 
i.e. by swapping the values of $\sigma_i(q)$ and $\sigma_i(r)$
for one or more indices $i \in [n]$. 
%
 Formally, the problem is 
 \begin{subequations}\label{2-assignment-exchange}
\begin{equation}
\min_{\sigma_1^{\ast},\ldots,\sigma_n^{\ast} \in \mathbf{S}_m} \sum_{1\le i<j\le n} \sum_{k=1}^m \left\| x_{i\sigma_i^{\ast}(k)} - x_{j\sigma_j^{\ast}(k)} \right\|^2 
\end{equation}
 under the constraints 
 \begin{equation}\label{2-assignment constraints}
\left\{ \begin{array}{l}
 \sigma_i^{\ast}(k) = \sigma_i(k) , \  k\in[m]\setminus \{k,l\} \\ 
 (\sigma_i^{\ast}(q) ,\sigma_i^{\ast}(r)) \in \{ (\sigma_i(q) ,\sigma_i(r)), (\sigma_i(r) ,\sigma_i(q)) \} 
\end{array} \right., \quad  1 \le i \le n .
\end{equation}
\end{subequations}

To fix ideas,  assume without loss of generality that $(q,r)=(1,2)$ and $\sigma_i = \mathrm{Id}_{[m]}$ for $1 \le i\le n$. 
Problem \eqref{2-assignment-exchange} becomes 
%
\begin{equation}\label{2-assignment simple}
\min_{\sigma^{\ast}_1,\ldots,\sigma^{\ast}_n \in \mathbf{S}_2} \sum_{1\le i,j \le n}  \big\| x_{i\sigma^{\ast}_i(1)} - x_{j\sigma^{\ast}_j(1)} \big\|^2 + \sum_{1\le i,j \le n}  \big\| x_{i\sigma^{\ast}_i(2)} - x_{j\sigma^{\ast}_j(2)} \big\|^2  \,.
\end{equation}

As in the previous sections, the problem can be transformed to 
$$\max_{\sigma^{\ast}_1,\ldots,\sigma^{\ast}_n \in \mathbf{S}_2}  \Big\| \sum_{i=1}^n   x_{i\sigma^{\ast}_i(1)} \Big\|^2 
+ \Big\| \sum_{i=1}^n   x_{i\sigma^{\ast}_i(2)}\Big\|^2   . $$
Replacing the permutations $\sigma^{\ast}_i \in \mathbf{S}_2$ by binary variables $c_i$, the problem becomes  
$$\max_{c_1,\ldots,c_n \in \{0,1\}}  \Big\| \sum_{i=1}^n  (c_i x_{i1} + (1-c_i) x_{i2}) \Big\|^2 
+ \Big\| \sum_{i=1}^n   ((1-c_i) x_{i1} + c_i x_{i2}) \Big\|^2  $$
and, after simple manipulations, 
\begin{equation}\label{UBQP}
\max_{c_1,\ldots,c_n \in \{0,1\}} \sum_{i,j}  c_i  c_j  \langle  d_i , d_j \rangle  
 - n  \sum_{i}  c_i  \langle d_i, \bar{d}\rangle       
\end{equation}
where $d_i = x_{i1}-x_{i2}$ and $\bar{d} = (1/n)\sum_{i=1}^n d_i$. This is an unconstrained binary quadratic program (UBQP) 
of size $n$ that can be solved with standard mathematical software (e.g. Cplex, Gurobi, Mosek).   
Refer to \cite{Kochenberger2014} for a survey of the UBQP literature. 



Having reduced
the basic pairwise interchange problem
\eqref{2-assignment-exchange} 
to the UBQP \eqref{UBQP},  
We now embed it in Algorithm 3.4 of \cite{Robertson2001} 
which combines randomization and greedy selection of interchange pairs. 
Hereafter $F(\sigma)$ denotes the objective value in \eqref{MDADC} and $\sigma= (\sigma_1,\ldots,\sigma_n) \in (\mathbf{S}_m)^n$
is identified with the assignments $\{ \sigma^1, \ldots , \sigma^m \}$, 
where $\sigma^l = (\sigma_1(l),\ldots , \sigma_n(l))$. 
The notation $\mathrm{diag}(\cdot)$ is used for diagonal matrices. 

\begin{algorithm}[ht!]
\caption{Pairwise Interchange with Greedy Selection}
\label{alg:2xchange}
\begin{algorithmic}[1]
\REQUIRE $X_1,\ldots, X_n \in \mathbb{R}^{p\times m}$, $\sigma \equiv \{ \sigma^{1}, \ldots, \sigma^{m}\}$
\STATE $\mathcal{C} \leftarrow  \sigma $
\COMMENT{candidate set of  assignments for interchange}
\WHILE {$\mathcal{C}  \ne \emptyset $}
\STATE $F_{best} \leftarrow F(\sigma) $
\STATE $\sigma^{+}\leftarrow \emptyset $, $\tau^{+} \leftarrow \emptyset $
\STATE Select $\sigma^q \in \mathcal{C}$
\FOR {$\sigma^r \in \mathcal{C} \setminus \{ \sigma^q \}$}
\STATE  $d_i \leftarrow x_{i \sigma^q(i)}-x_{i \sigma^{r}(i)} \ (1\le i \le n)$, $\bar{d} \leftarrow \frac{1}{n} \sum_{i=1}^n d_i$
\STATE $ Q \leftarrow (\langle d_i ,d_j \rangle)_{1 \le i, j \le m} - \mathrm{diag}(n\langle d_1,\bar{d}\rangle, \ldots, n
 \langle d_1,\bar{d}\rangle ) $
\STATE Solve the UBQP \eqref{UBQP} with quadratic matrix $Q$ and 
call $ (c_1,\ldots,c_n )$ a solution.  
\STATE $\tilde{\sigma}^{q}(i) \leftarrow c_i \, \sigma^{q}(i) + (1 -c_i)\, \sigma^{r}(i) \, (1 \le i \le n)$
\STATE $\tilde{\sigma}^{r}(i) \leftarrow c_i \, \sigma^{r}(i) + (1-c_i)\, \sigma^{q}(i) \, (1 \le i \le n)$
\STATE $\tilde{F} \leftarrow F( \sigma \setminus \{ \sigma^q,\sigma^r \} \cup \{ \tilde{\sigma}^q ,\tilde{\sigma}^r\})$ 
\IF {$\tilde{F} < F_{best}$} 
\STATE $(\sigma^{+} ,\tau^{+})\leftarrow (\tilde{\sigma}^q ,  \tilde{\sigma}^r)$
\COMMENT{candidate new pair of assignments}
\STATE ($\sigma^{-},\tau^{-}) \leftarrow (\sigma^q, \sigma^r)$
\COMMENT{candidate old pair of assignments}
\STATE $F_{best} \leftarrow \tilde{F} $
\ENDIF 
\ENDFOR
\IF{$\sigma^{+} \ne \emptyset$}
\STATE $\sigma \leftarrow \sigma \setminus \{ \sigma^{-} , \tau^{-} \} \cup \{ \sigma^{+}, \tau^{+}\} $
\COMMENT{perform interchange}
\STATE $\mathcal{C} \leftarrow \sigma $
\COMMENT{reset candidate set to all assignments}
\ELSE 
\STATE  $\mathcal{C} \leftarrow  \mathcal{C} \setminus \{ \sigma^q\}$
\COMMENT{remove assignment from candidate set}
\ENDIF 
\ENDWHILE
\end{algorithmic}
\end{algorithm}

\subsection{Gaussian mixture approach}
\label{sec: GM}

The matching problem \eqref{MDADC} has a probabilistic interpretation in terms of mixture models.  
Let $y_1 , \ldots, y_m$ be random vectors in $\mathbb{R}^p$ 
with respective probability distributions $\mathcal{P}_{1},\ldots, \mathcal{P}_m$. 
Assume that these vectors  are only observable 
 after their labels have been shuffled at random. 
The random permutation of labels 
represents the uncertainty about the correspondence between observations, 
say  $x_1,\ldots, x_m $,  
and their underlying distributions $\mathcal{P}_{1}, \ldots, \mathcal{P}_{m}$.   
For mathematical convenience, $y_1 , \ldots, y_m$ 
are assumed independent and each $\mathcal{P}_{k}$  $(1\le k \le m )$ 
is taken as a multivariate normal distribution $  N(\mu_k,\Sigma_k)$. 
The data-generating process can be summarized as
\begin{equation}\label{GM}
\left\{
\begin{array}{l}
 y_k \sim N(\mu_k, \Sigma_k) \quad (1\le k \le m),  \\
s \textrm{ has a uniform distribution over } \mathbf{S}_m ,  \\
 (y_1,\ldots,y_m ) \textrm{ are mutually independent and independent of } s , \\
 (x_1, \ldots, x_m) = (y_{s(1)}, \ldots, y_{s(m)}) .
 \end{array} \right.
\end{equation}
This can be viewed as a Gaussian mixture model 
with permutation constraints on cluster assignments.  
These constraints can be shifted to the mean 
and covariance parameters by concatenating observations: 
the vector $x =\mathrm{vec} (x_1,\ldots,x_m)$ follows a
mixture of $m!$ multivariate normal distributions $N(\mu_{\sigma}, \Sigma_{\sigma})$ 
in $\mathbb{R}^{mp}$ with equal mixture weights $1/m!$,  
where  $\mu_{\sigma} = \mathrm{vec}(\mu_{\sigma(1)}, \ldots,\mu_{\sigma(m)})$ and 
$\Sigma_{\sigma} = \mathrm{diag} (\Sigma_{\sigma(1)},\ldots, \Sigma_{\sigma(m)})$ (block-diagonal matrix) 
for $\sigma  \in \mathbf{S}_{m}$;  see also \cite{Qiao2015}.
In this form, the theory and methods of 
Gaussian mixture models are seen to apply to \eqref{GM},
in particular the consistency and asymptotic normality of maximum likelihood estimators 
\citep[Chapter 2]{McLachlan2000}. 
\smallskip
\begin{remark}
In model \eqref{GM}, the cluster centers 
  $\{\overline{x}_{\hat{\sigma}, 1},\ldots,  \overline{x}_{\hat{\sigma}, m}\} $ 
  associated to a global solution $\hat{\sigma} = (\hat{\sigma}_1,\ldots,\hat{\sigma}_n)$ of 
 problem \eqref{MDADC} are \emph{not} consistent for $\{\mu_1,\ldots, \mu_m\}$ as $n\to\infty$. 
Consider for example the case where $p=1$,  $m=2$ (univariate mixture with two components), 
and  $\mu_1 < \mu_2$. Then $\hat{\mu}_1 = \frac{1}{n}\sum_{i=1}^n \min(x_{i1},x_{i2})$ 
and  $\hat{\mu}_2 = \frac{1}{n}\sum_{i=1}^n \max(x_{i1},x_{i2})$. Accordingly $E(\hat{\mu}_1) = E(\min(x_1,x_2)) < \mu_1$ 
and $E(\hat{\mu}_2) = E(\max(x_1,x_2)) > \mu_2$, meaning that both estimators are biased and inconsistent.  
\end{remark}
\smallskip

\begin{remark}
The permutation constraints of model \eqref{GM}
can be formulated as \emph{equivalence constraints}  \citep[see][and Section \ref{sec:introduction}]{Shental2003}. 
However, this general formulation is unlikely to lead to faster or better optimization, 
just as the constrained $K$-means approach of \cite{Wagstaff2001}, 
which also handles equivalence constraints, 
does not improve upon the specialized $K$-means Algorithm \ref{alg:kmeanslike} 
for problem \eqref{MDADC} (see section \ref{sec:simulations}). 
\end{remark}

Gaussian mixture models and the 
 Expectation Maximization (EM) algorithm 
\citep[see e.g.][]{McLachlan2000,McLachlan2008} 
constitute a well-known approach to clustering. 
Here, in view of the matching problem \eqref{MDADC}, 
we propose a computationally efficient EM approach to 
the Gaussian mixture model \eqref{GM}.
Although in principle, the standard EM 
algorithm for a Gaussian 
mixture model could be applied,  
the number $m!$ of mixture components 
and the potentially high dimension $mp$ 
of the data in \eqref{GM} 
render computations intractable 
 unless $m$ is very small.

Let $(x_{i1},\ldots,x_{im})$ ($1\le i \le n$) be data arising from 
 \eqref{GM} and let $s_1, \ldots, s_n$ be associated label 
permutations. For convenience, the permutations are expressed in terms of indicator variables
$I_{ikl}$ ($1\le i \le n, \ 1\le k,l \le m$): 
$I_{ikl}=1$ if $x_{ik} = y_{il}$ or equivalently $s_i(k)=l$, $I_{ikl}=0$ otherwise. 
The $(x_{ik})$ and  $(I_{ikl})$ are the so-called complete data.  
 Call $\hat{\theta} = \{ (\hat{\mu}_{l}, \hat{\Sigma}_{l}) : l\in[ m]\}$ 
 the current estimate of the model parameters of \eqref{GM} 
 in the EM procedure. 
The  log-likelihood of the complete data is 
\begin{equation}\label{loglik complete}
 \log L_c = \sum_{i=1}^{n} \sum_{k=1}^m \sum_{l=1}^{m} \log \varphi (x_{ik} ; \hat{\mu}_l, \hat{\Sigma}_l) I_{ikl} 
\end{equation}
where $\varphi(x;\mu,\Sigma) = (2\pi)^{-p/2} | \Sigma |^{-1/2} \exp\big( - (x-\mu)'\Sigma^{-1}(x-\mu) / 2\big)$ 
indicates a multivariate normal density in $\mathbb{R}^p$. 

\paragraph{E step.}
The E step of the EM algorithm consists in calculating the expected value of 
\eqref{loglik complete} conditional on the observed data $X_1,\ldots,X_n$ 
and assuming that $\theta = \hat{\theta}$. This amounts to deriving, 
for each $(i,k,l)$, the quantity 
\begin{align}
E_{\hat{\theta}}(  I_{ikl} | X_i) & = P_{\hat{\theta}} ( I_{ikl} = 1| X_i) \nonumber \\
& = \frac{ P_{\hat{\theta}} (X_i | I_{ikl} = 1) P_{\hat{\theta}}( I_{ikl} = 1) } { P_{\hat{\theta}} (X_i) } \nonumber\\
& = c_i  P_{\hat{\theta}} (X_i | I_{ikl} = 1)\nonumber \\ 
& = c_i  \sum_{\sigma \in \mathbf{S}_m : \sigma(k)=l} P_{\hat{\theta}}\big(X_i | I_{i1 \sigma(1)} = 1,\ldots,I_{im \sigma(m)} = 1\big) \nonumber \\
& \qquad \qquad  \qquad \times
P_{\hat{\theta}}\big( I_{i1 \sigma(1)} = 1,\ldots,I_{im \sigma(m)} = 1 \big| I_{ik l} = 1\big)\nonumber \\
& = \frac{c_i}{(m-1)!} \sum_{\sigma \in \mathbf{S}_m : \sigma(k)=l}  \prod_{r=1}^{m} P_{\hat{\theta}} (x_{ir} | I_{ir \sigma(r)} = 1)\nonumber \\
& = \frac{c_i}{(m-1)!} \sum_{\sigma \in \mathbf{S}_m : \sigma(k)=l}  \prod_{r=1}^{m} \varphi\big( x_{ir} ; \hat{\mu}_{ \sigma(r)} , \hat{\Sigma}_{\sigma(r)}\big)\,. \label{exact E step}
\end{align}

Formula \eqref{exact E step} can be conveniently expressed with \emph{matrix permanents}. 
The permanent of a square matrix $A=(a_{ij})$ of dimension $m\times m$ is defined as 
$\mathrm{per}(A) = \sum_{\sigma \in \mathbf{S}_m} \prod_{i=1}^{m} a_{i\sigma (i)}$. 
Writing $A_i = (a_{ikl}) =  (\varphi ( x_{ik} ; \hat{\mu}_{l} , \hat{\Sigma}_{l} )) \in \mathbb{R}^{m\times m}$ and  
$A_i^{-(k,l)} = (a_{ik'l'}  )_{k'\ne k, l'\ne l}\in\mathbb{R}^{(m-1)\times (m-1)} $, 
\eqref{exact E step} reformulates as 
$ E_{\hat{\theta}}(  I_{ikl} | X_i)  = a_{ikl}\, \mathrm{per}(A_i^{-(k,l)}) / \mathrm{per}(A_i) $.

The permanent of a matrix has a very similar expression to the Leibniz formula for determinants, 
but without the  permutation signatures $\pm 1$.  
It is however far more expensive to compute: 
 efficient implementations 
have complexity $O(2^m m^2)$ \citep{Ryser1963} 
or $O(2^m m)$ \citep{Nijenhuis1978}. 
Stochastic approximation methods running in polynomial time
\citep[e.g.][]{Jerrum2004,Kuck2019} 
and variational bounds 
\citep[see][and the references therein]{Uhlmann2004} are also available. 
Given that \eqref{exact E step} must be evaluated for $nm^2$ values of $(i,k,l)$, 
and accounting for the computation of the matrices $A_i $ ($1\le i \le n$)  
  \citep[e.g.][Chap. 16.1]{Press2007}, 
the E step has overall complexity at least $O(2^m m^3 n+ mp^3 + m^2 p^2 n)$. 

The evaluation of permanents requires precautions to avoid numerical underflow. 
Indeed, the density values $\varphi ( x_{ik} ; \hat{\mu}_{l} , \hat{\Sigma}_{l} ) $ are often very small 
and multiplying them in \eqref{exact E step} may quickly lead to numerical zeros. 
Preconditioning greatly helps in this regard:
by the properties of the permanent, multiplying the rows and columns of $A_i $ by nonzero numbers has no effect on 
\eqref{exact E step} as these multiples cancel out between the numerator $a_{ikl} \mathrm{per}(A_i^{-(k,l)})$ and 
denominator $\mathrm{per}(A_i)$. One can exploit this by alternatively rescaling the rows and columns of $A_i$ by their sums. 
Provided that $A_i$ is a positive matrix, 
this scheme 
converges to a doubly stochastic matrix \citep{Sinkhorn1964} 
that in practice often has at least one ``non-small" entry in each row and each column.

\paragraph{M step.} By standard least square calculations, the updated estimate 
 $\theta^{+}  = \{ (\mu_{l}^{+} , \Sigma_{l}^{+}) : 1 \le l \le m \} $ is 
\begin{equation}\label{M step}
\begin{split} 
\mu_{l}^{+} & = \frac{1}{n} \sum_{i=1}^n \sum_{k=1}^{m} P_{\hat{\theta}}(I_{ikl} = 1| X_i) x_{ik} \\
   \Sigma_{l}^{+} & =  \frac{1}{n} \sum_{i=1}^n \sum_{k=1}^{m} P_{\hat{\theta}}(I_{ikl} = 1|X_i) (x_{ik} - \mu_{l}^{+})(x_{ik} - \mu_{l}^{+})' 
\end{split}
\end{equation}
with  $P_{\hat{\theta}}(I_{ikl} = 1 | X_{i}) = E_{\hat{\theta}}(  I_{ikl} | X_{i}) $ given by \eqref{exact E step}.  
The fact that $ \sum_{k=1}^{m} P_{\hat{\theta}}(I_{ikl} = 1|X_i) = 1$ for all $(i,l)$ was used to simplify \eqref{M step}.
If the variances $\Sigma_{1},\ldots,\Sigma_{m}$ are assumed equal, their common estimate should be 
$\Sigma^{+} = (1/m) \sum_{l=1}^m \Sigma_{l}^{+}$.

\paragraph{Log-likelihood.} The log-likelihood of the observed data is given by 
\begin{equation}\label{loglik}
 \log L(\hat{\theta})  = \sum_{i=1}^{n}  \log\left( \frac{1}{m!}  \sum_{\sigma \in S}  \prod_{k=1}^{m} \varphi\big( x_{ik} ; \hat{\mu}_{ \sigma(k)} , \hat{\Sigma}_{\sigma(k)}\big) \right) .
\end{equation}
It is simply the sum of the logarithms of the permanents of the matrices 
$A_i =\big( \varphi ( x_{ik} ; \hat{\mu}_{ l} , \hat{\Sigma}_{l}) \big)$ 
defined earlier.
Since these permanents are calculated in the E step, there is essentially no additional 
 cost to computing the log-likelihood.

The implementation of the EM algorithm for model \eqref{GM} is sketched in Algorithm \ref{alg:EM}, 
The initial covariance matrices $\Sigma_1, \ldots, \Sigma_m$ in this algorithm 
should be taken positive definite to avoid degeneracy issues when evaluating multivariate normal densities. 
However, the algorithm is easily extended to handle singular covariance matrices.  
In practice, stopping criteria for the EM algorithm are often based on the absolute or relative increase in 
log-likelihood between successive iterations.

\begin{algorithm}[ht!]
\caption{EM for Constrained Gaussian Mixture} 
\label{alg:EM}
\begin{algorithmic}[1]
\REQUIRE $X_1,\ldots, X_n \in \mathbb{R}^{p\times m}$, $\mu_1, \ldots, \mu_m \in \mathbb{R}^p$, 
$ \Sigma_{1}, \ldots, \Sigma_{m} \in \mathbb{R}^{p\times p}$
\STATE $\theta^{0} \leftarrow  \{ (\mu_l ,\Sigma_l) : 1\le l \le m\} $
\FOR {$t=0,1,\ldots$}  
\STATE Perform Choleski decomposition $\Sigma_l = L_l'L_l$ 
with $L_l$ lower triangular ($ 1 \le l \le m$)
\FOR{$i=1,\ldots , n$}
\STATE  $a_{ikl} \leftarrow (2\pi)^{-p/2} | L_l |^{-1/2} e^{  - \| L_l^{-1}(x_{ik}-\mu_l )\|^2 / 2} \ \  (1\le k,l \le m)$,  
   $A_i \leftarrow (a_{ikl})$
\FOR{$k=1,\ldots, m$}
\FOR{$l=1,\ldots,m$}
\STATE Alternatively rescale rows and columns
 of $A_{i}^{-(k,l)}$  to sum to 1 
 \STATE Calculate $\mathrm{per}(A_{i}^{-(k,l)})$ with Ryser's inclusion-exclusion formula
%
 \STATE $p_{ikl} \leftarrow  a_{ikl}  \, \mathrm{per}(A_{i}^{-(k,l)}) $ 
 \ENDFOR 
\ENDFOR 
\STATE $c_i \leftarrow  \frac{1}{m} \sum_{k=1}^m \sum_{l=1}^{m} p_{ikl} $
 \STATE $ w_{ikl} \leftarrow   p_{ikl}  / c_i  \ 
 (1\le k,l \le m)$ 
\COMMENT{class membership probability}
 \ENDFOR 
 \STATE $\ell^{t} \leftarrow \sum_{i=1}^n \log c_i$
 \COMMENT{log-likelihood} 
 \FOR{$l = 1,\ldots, m$}
 \STATE $\mu_{l} \leftarrow  \frac{1}{n} \sum_{i=1}^n \sum_{k=1}^{m} w_{ikl} x_{ik} $
\STATE $\Sigma_{l} \leftarrow \frac{1}{n} \sum_{i=1}^n \sum_{k=1}^{m}w_{ikl}  (x_{ik} - \mu_{l})(x_{ik} - \mu_{l})'  $
 \ENDFOR
 \STATE $\theta^{t+1} \leftarrow  \{ (\mu_l ,\Sigma_l) : 1\le l \le m\} $
 \ENDFOR 
\end{algorithmic}
\end{algorithm}

In statistical problems involving a large number of latent variables such as \eqref{GM}, 
the EM algorithm is usefully extended by the so-called \emph{deterministic annealing EM} algorithm \citep[DAEM,][]{Ueda1998}. The DAEM is identical to the EM  except that in the E step, 
the assignment probabilities $ P_{\theta} ( I_{ikl} = 1| X_i)$ are raised to a power $\beta \in (0,1]$ 
and rescaled to remain valid probabilities. This effectively flattens out the differences 
between assignment probabilities, keeping the uncertainty about cluster/class assignment relatively high. 
As the number $t$ of iterations grows, the power $\beta = \beta_t$, which represents an inverse temperature parameter, 
increases to 1. For $t$ sufficiently large, 
the DAEM reverts back to the EM. 
In this way the DAEM offers some control 
on how many iterations are spent exploring the latent variable space 
before converging to a set of (often highly unbalanced) assignment probabilities. 
In particular, appropriate 
use of the DAEM prevents the convergence from happening too fast.

%
%
%
%

\subsection{Algorithm initialization} 

The matching methods developed for \eqref{MDADC} in the previous sections are local search procedures.  
As can be expected, the quality of their solutions largely depends on their starting points. 
Several strategies for finding good starting points are presented hereafter.

\emph{Random initialization.} 
 Utilizing multiple random starting points $\sigma \in (\mathbf{S}_m)^n$ 
 or $P \in (\Pi_m)^n$ often yields at least one nearly optimal solution. 
 This strategy is particularly suitable when the computational cost of optimization is cheap, 
 as is the case with Algorithms \ref{alg:kmeanslike}-\ref{alg:bca}-\ref{alg:Frank-Wolfe}.   


\emph{Template matching.} 
Given data matrices $X_1,\ldots, X_m \in \mathbb{R}^{p\times m}$ and a template matrix 
$ T \in \mathbb{R}^{p\times m} $, solve the matching problem
 \begin{equation}\label{template} 
\min_{P_1,\ldots, P_n \in \Pi_m } \sum_{i=1}^{n} \left\| X_i P_i - T \right\|_F^2  .
\end{equation}
The expediency of template matching comes from the fact that it reduces ${n \choose 2}$ related matching tasks between pairs of data matrices in \eqref{MDADC} to $n$ separate matching tasks between the data and the template. 
A central question is: which template to use? 
\cite{Bandelt1994} propose to  either take a single data matrix 
as template (\emph{single hub heuristic}), e.g. $T=X_1$, 
or to examine all data matrices in turn: $T\in\{ X_1,\ldots, X_n\}$,  
and retain the assignment  $ P(T) = (P_1(T), \ldots, P_n(T))$ 
that yields the lowest value of \eqref{MDADC} (\emph{multiple hub heuristic}). 
More generally, the template need not be a data point; it could for example be an estimate of cluster centers based on previous observations. 
%

\emph{Recursive heuristics.} 
The recursive heuristics of \cite{Bandelt1994} (see Section \ref{sec:introduction}) 
are easily applicable to problem \eqref{MDADC}. 
Their algorithm RECUR1 for example,  
which is related to the BCA Algorithm \ref{alg:bca}, is implemented as follows. 
The first permutation matrix $P_1$ can be selected arbitrarily, say $P_1 = I_{m}$. 
Then for $i=1,\ldots, n-1$, the LAP \eqref{LAP} is changed to\break   
\begin{equation}
 \max_{P_{i+1} \in \Pi_{m} }  \Big\langle P_{i+1}\, ,\,   X_{i+1}' \sum_{ j=1  }^{i} X_j P_j \Big\rangle_F \,.
\end{equation}



\section{Numerical study} 
\label{sec:simulations}

This section presents experiments that assess the numerical and computational performances of the matching methods of Section \ref{sec:methods} and other relevant methods from the literature.
Three performance measures are reported: the attained objective value in the matching problem \eqref{MDADC}, the Rand index \citep{Rand1971} for evaluating agreement between matchings and data labels, and the computation time.

\paragraph{Simulation setup.} 

The simulations are based on handwritten digits data available on the UCI machine learning repository 
(\href{https://archive.ics.uci.edu/ml/datasets/Optical+Recognition+of+Handwritten+Digits}{archive.ics.uci.edu}). 
Unlike classification problems, 
the task at hand is to match collections of digits without using label information. 
The data are normalized bitmaps of handwritten digits. After downsampling, images of dimensions $8\times 8$ are obtained with integer elements in $\{0, \ldots, 16\}$. The training data used for the simulations contain 3823 images contributed by 30 people, with about  380 examples for each digit $0,\ldots,  9$. 
A principal component analysis (PCA) is carried out separately for each of the $m=10$ digit classes (after vectorizing 
the $8\times8$ input matrices) and the 25 first PCs are retained for each class, which represents at least 95\% of the class variance. 
Artificial data are then generated according to the model  $x_{ik} = \sum_{r=1}^{25} \xi_{ikr} \phi_{kr} + \varepsilon_{ikr}$ 
for $ 1\le i \le n$ and $1\le k \le m$, where the $\phi_{kr} $ are PC vectors of length $p=64$ 
and the $\xi_{ikr}$ are independent normal random variables with mean zero 
and standard deviation given by the PCA. 
A small amount of Gaussian white noise $\varepsilon$ with standard deviation 2.5 is added to the simulated data, 
which corresponds to 10\% of the standard deviation of the original data. The number $n$ of statistical units varies in 
$\{ 5, 10 , 20 , 30, 40 , 50, 75, 100, 200, 500, 1000\}$. For each value of $n$, 
the simulation is replicated 100 times. The simulations are run in the R programming environment. 
Code for the simulations and  the R package \texttt{matchFeat} implementing all methods of this paper are available at  
\href{https://github.com/ddegras/matchFeat}{github.com/ddegras/matchFeat}.

\paragraph{Matching methods.}
The methods of Section \ref{sec:methods} are combined in three steps: 
initialization, main algorithm, and optional post-processing. 
Four  initializations are considered: identity matrix (ID), 100 random starting points (R100), 
multiple-hub heuristic (HUB), and recursive heuristic (REC). 
A fifth initialization clustering data vectors by their digit labels (LBL) 
is also examined as a benchmark. This initialization is infeasible in practice;  
 it may also not minimize \eqref{MDADC} although it is often nearly optimal. 
The main algorithms are $K$-means matching (KM), block coordinate ascent (BCA), and 
the Frank-Wolfe method (FW).  
The pairwise interchange algorithm (2X) and EM algorithm for constrained Gaussian mixture (EM) are used for post-processing only as they were seen to perform poorly on their own (that is, with any of the proposed initializations) in 
preliminary experiments. 
The simulations also comprise matching methods representative of the literature:  

\begin{itemize}

\item[-]  \emph{Integer linear program (ILP).} The standard ILP formulation of the MDADC \eqref{MAP}-\eqref{DC} \citep[e.g.][]{Kuroki2009,Tauer2013} involves ${n \choose 2} m^2$ binary variables (the number of edges in a complete $n$-partite graph with $m$ nodes in each subgraph), $n(n-1)m$ equality constraints and ${n \choose 3}m^3$ inequality constraints (so-called triangle or clique constraints). Another formulation of the ILP  expresses the triangle constraints with reference to one the $n$ subgraphs, 
thereby reducing their number to ${n \choose 2} m^3$. 
 
\item[-] \emph{ILP relaxation and integer quadratic program (IQP).}   
Two of the methods in \cite{Kuroki2009} are considered: the first consists in dropping the triangle constraints, 
solving ${n \choose 2 }$ separate assignment problems, and recovering a proper solution with multiple-hub heuristics. 
The second expresses the triangle constraints with reference to one of the $n$ subgraphs as in the above ILP, 
and formulates the objective function only in terms of those edges starting from and arriving to the reference subgraph. 
This reduces the number of optimization variables to ${n \choose 2}m^2$ but transforms the linear program into a quadratic one.


\item[-] \emph{Constrained $K$-means.} The  
COP-KMEANS  \citep{Wagstaff2001}, MPCK-MEANS \citep{Bilenko2004},
  LCVQE \citep{Pelleg2007}, 
 and CCLS \cite{Hiep2016} 
 algorithms all handle equivalence constraints and can thus be applied to \eqref{MDADC}. 
 They are conveniently implemented in the R package \texttt{conclust} of the last authors. 
COP-KMEANS and CCLS treat equivalence constraints as hard constraints and thus exactly solve \eqref{MDADC}. 
MPCK-MEANS and LCVQE handle equivalence constraints as soft constraints (in addition, MPCK-MEANS 
incorporates metric learning) and thus approximately solve \eqref{MDADC}. 

\end{itemize}
Going forward, these methods will be referred to as 
ILP, KUR-ILP, KUR-IQP, COP-KM, MPC-KM, LCVQE, and CCLS. 
While they are applicable to the sum-of-squares matching problem \eqref{MDADC}, 
these methods are not geared towards it and should not be expected to outperform the methods of this paper. 
Lagrangian heuristics \citep[e.g.][]{Tauer2013,Natu2020} are not included in the simulations because  their efficient implementation requires computer clusters and/or specialized computing architecture, whereas the focus of this paper is on methods executable on a single machine. 
\smallskip

\begin{remark}
Initial attempts were made to obtain lower bounds on the global minimum in \eqref{MDADC} using a relaxation method of \cite{Bandelt2004}. However, the resulting bounds are far too small, 
a fact already noted by these authors  in the case of non-Euclidean distances $d$ (recall that in \eqref{MDADC}, 
 $d$ is the \emph{squared} Euclidean distance). 
\end{remark}

\paragraph{Results.}

\emph{Optimization accuracy.} 
%
To facilitate comparisons, we discuss the relative error of each method 
averaged across 100 replications for each $n$. The relative error of a method 
is defined as the ratio of its attained objective value in \eqref{MDADC}
by the minimum objective value across all methods minus 1.  
Full results are available in Table \ref{table:optimization}. 
Hereafter and in the table, methods are listed by order of best performance.

\begin{table}[ht!]
\hspace*{-15mm}
\scalebox{.6}{
\begin{tabular}{r *{11}{c} }
  \hline
Method & $n=5$ & $n=10$ & $n=20$ & $n=30$ & $n=40$ & $n=50$ & $n=75$ & $n=100$ & $n=200$ & $n=500$ & $n=1000$ \\ 
  \hline
R100-BCA &2E-11 (1E-10)& 0 (0)& 0 (0)& 0 (0)& 0 (0)& 0 (0)& 0 (0)& 0 (0)& 0 (0)& 0 (0)& 0 (0)\\
R100-BCA-2X &2E-11 (1E-10)& 0 (0)& 0 (0)& 0 (0)& 0 (0)& 0 (0)& 0 (0)& 0 (0)&  &  &  \\
KUR-IQP &2E-11 (1E-10)&  &  &  &  &  &  &  &  &  &  \\
ILP &0 (0)& 3E-5 (3E-4)&  &  &  &  &  &  &  &  &  \\
LBL-BCA &2E-3 (4E-3)& 1E-3 (3E-3)& 4E-4 (1E-3)& 3E-4 (1E-3)& 1E-4 (4E-4)& 2E-4 (6E-4)& 5E-5 (3E-4)& 2E-5 (1E-4)& 3E-6 (3E-5)& 3E-8 (1E-7)& 1E-8 (6E-8)\\
LBL-BCA-2X &2E-3 (3E-3)& 8E-4 (2E-3)& 2E-4 (5E-4)& 1E-4 (4E-4)& 6E-5 (2E-4)& 1E-4 (4E-4)& 4E-5 (2E-4)&  &  &  &  \\
HUB-BCA-2X &1E-3 (2E-3)& 1E-3 (3E-3)& 4E-4 (1E-3)& 2E-4 (1E-3)& 2E-4 (6E-4)& 3E-4 (9E-4)& 6E-5 (3E-4)&  &  &  &  \\
HUB-BCA &1E-3 (3E-3)& 2E-3 (3E-3)& 8E-4 (2E-3)& 3E-4 (1E-3)& 5E-4 (2E-3)& 5E-4 (1E-3)& 2E-4 (1E-3)& 1E-4 (7E-4)& 1E-4 (6E-4)& 2E-5 (2E-4)& 1E-8 (5E-8)\\
LBL-FW-2X &4E-3 (6E-3)& 2E-3 (3E-3)& 5E-4 (9E-4)& 2E-4 (4E-4)& 2E-4 (4E-4)& 2E-4 (4E-4)& 9E-5 (2E-4)&  &  &  &  \\
REC-BCA-2X &3E-3 (5E-3)& 2E-3 (5E-3)& 8E-4 (3E-3)& 6E-4 (2E-3)& 3E-4 (8E-4)& 2E-4 (7E-4)& 8E-5 (3E-4)&  &  &  &  \\
LBL-KM-2X &4E-3 (5E-3)& 2E-3 (3E-3)& 5E-4 (9E-4)& 2E-4 (4E-4)& 2E-4 (4E-4)& 2E-4 (4E-4)& 9E-5 (2E-4)&  &  &  &  \\
ID-BCA-2X &3E-3 (6E-3)& 2E-3 (3E-3)& 1E-3 (3E-3)& 6E-4 (2E-3)& 4E-4 (1E-3)& 4E-4 (1E-3)& 1E-4 (5E-4)&  &  &  &  \\
R100-FW-2X &7E-3 (9E-3)& 3E-3 (4E-3)& 2E-4 (5E-4)& 4E-5 (1E-4)& 2E-5 (7E-5)& 3E-6 (1E-5)& 3E-6 (2E-5)& 2E-6 (6E-6)&  &  &  \\
REC-BCA &4E-3 (7E-3)& 4E-3 (8E-3)& 1E-3 (3E-3)& 1E-3 (3E-3)& 8E-4 (2E-3)& 6E-4 (2E-3)& 5E-4 (2E-3)& 1E-4 (5E-4)& 2E-4 (8E-4)& 3E-4 (1E-3)& 7E-5 (7E-4)\\
R100-KM-2X &9E-3 (1E-2)& 3E-3 (4E-3)& 2E-4 (5E-4)& 4E-5 (1E-4)& 2E-5 (7E-5)& 3E-6 (1E-5)& 3E-6 (2E-5)& 2E-6 (6E-6)&  &  &  \\
ID-BCA &5E-3 (9E-3)& 5E-3 (8E-3)& 3E-3 (6E-3)& 2E-3 (4E-3)& 8E-4 (2E-3)& 9E-4 (2E-3)& 5E-4 (2E-3)& 6E-4 (1E-3)& 3E-4 (1E-3)& 4E-4 (1E-3)& 1E-8 (5E-8)\\
HUB-FW-2X &4E-3 (6E-3)& 4E-3 (6E-3)& 1E-3 (1E-3)& 8E-4 (1E-3)& 6E-4 (8E-4)& 4E-4 (8E-4)& 2E-4 (4E-4)&  &  &  &  \\
HUB-KM-2X &5E-3 (6E-3)& 5E-3 (6E-3)& 1E-3 (2E-3)& 8E-4 (1E-3)& 6E-4 (8E-4)& 4E-4 (8E-4)& 2E-4 (4E-4)&  &  &  &  \\
REC-KM-2X &6E-3 (9E-3)& 5E-3 (6E-3)& 2E-3 (4E-3)& 1E-3 (3E-3)& 9E-4 (2E-3)& 4E-4 (1E-3)& 2E-4 (5E-4)&  &  &  &  \\
REC-FW-2X &6E-3 (8E-3)& 5E-3 (6E-3)& 3E-3 (6E-3)& 1E-3 (3E-3)& 9E-4 (2E-3)& 4E-4 (1E-3)& 2E-4 (5E-4)&  &  &  &  \\
LBL-FW &2E-2 (2E-2)& 8E-3 (7E-3)& 2E-3 (2E-3)& 1E-3 (1E-3)& 7E-4 (7E-4)& 5E-4 (8E-4)& 3E-4 (5E-4)& 9E-5 (1E-4)& 2E-5 (4E-5)& 3E-6 (4E-6)& 8E-7 (7E-7)\\
LBL-KM &2E-2 (2E-2)& 8E-3 (7E-3)& 2E-3 (2E-3)& 1E-3 (1E-3)& 7E-4 (7E-4)& 5E-4 (8E-4)& 3E-4 (5E-4)& 9E-5 (1E-4)& 2E-5 (4E-5)& 3E-6 (4E-6)& 8E-7 (7E-7)\\
ID-KM-2X &9E-3 (1E-2)& 7E-3 (9E-3)& 4E-3 (8E-3)& 2E-3 (5E-3)& 2E-3 (5E-3)& 1E-3 (3E-3)& 7E-4 (3E-3)&  &  &  &  \\
ID-FW-2X &1E-2 (1E-2)& 6E-3 (8E-3)& 4E-3 (8E-3)& 2E-3 (3E-3)& 1E-3 (3E-3)& 1E-3 (4E-3)& 8E-4 (3E-3)&  &  &  &  \\
LBL &2E-2 (2E-2)& 1E-2 (9E-3)& 5E-3 (3E-3)& 3E-3 (2E-3)& 3E-3 (2E-3)& 3E-3 (2E-3)& 2E-3 (1E-3)& 2E-3 (9E-4)& 2E-3 (6E-4)& 2E-3 (4E-4)& 2E-3 (3E-4)\\
HUB-KM &3E-2 (1E-2)& 2E-2 (1E-2)& 6E-3 (5E-3)& 3E-3 (3E-3)& 2E-3 (3E-3)& 1E-3 (2E-3)& 9E-4 (2E-3)& 3E-4 (9E-4)& 2E-4 (7E-4)& 2E-5 (2E-4)& 8E-7 (8E-7)\\
HUB-FW &3E-2 (1E-2)& 2E-2 (1E-2)& 6E-3 (5E-3)& 3E-3 (3E-3)& 2E-3 (3E-3)& 1E-3 (2E-3)& 9E-4 (2E-3)& 3E-4 (9E-4)& 2E-4 (7E-4)& 2E-5 (2E-4)& 8E-7 (8E-7)\\
REC-KM &2E-2 (2E-2)& 2E-2 (1E-2)& 1E-2 (1E-2)& 5E-3 (6E-3)& 3E-3 (4E-3)& 3E-3 (6E-3)& 1E-3 (3E-3)& 9E-4 (3E-3)& 5E-4 (1E-3)& 5E-4 (2E-3)& 1E-4 (9E-4)\\
REC-FW &2E-2 (2E-2)& 2E-2 (1E-2)& 1E-2 (1E-2)& 5E-3 (6E-3)& 3E-3 (4E-3)& 3E-3 (6E-3)& 1E-3 (3E-3)& 9E-4 (3E-3)& 5E-4 (1E-3)& 5E-4 (2E-3)& 1E-4 (9E-4)\\
2X &1E-2 (1E-2)& 7E-3 (7E-3)& 5E-3 (5E-3)& 4E-3 (4E-3)&  &  &  &  &  &  &  \\
R100-FW &9E-2 (3E-2)& 1E-2 (7E-3)& 7E-4 (1E-3)& 1E-4 (2E-4)& 8E-5 (1E-4)& 3E-5 (5E-5)& 1E-5 (3E-5)& 7E-6 (1E-5)& 2E-6 (4E-6)& 3E-7 (5E-7)& 1E-7 (2E-7)\\
R100-KM &9E-2 (3E-2)& 1E-2 (7E-3)& 7E-4 (1E-3)& 1E-4 (2E-4)& 8E-5 (1E-4)& 3E-5 (5E-5)& 1E-5 (3E-5)& 7E-6 (1E-5)& 2E-6 (4E-6)& 3E-7 (5E-7)& 1E-7 (2E-7)\\
REC &2E-2 (2E-2)& 2E-2 (2E-2)& 2E-2 (2E-2)& 2E-2 (1E-2)& 2E-2 (1E-2)& 2E-2 (1E-2)& 2E-2 (1E-2)& 1E-2 (1E-2)& 9E-3 (8E-3)& 5E-3 (5E-3)& 4E-3 (4E-3)\\
ID-KM &3E-1 (9E-2)& 9E-2 (5E-2)& 3E-2 (2E-2)& 1E-2 (1E-2)& 5E-3 (9E-3)& 5E-3 (9E-3)& 3E-3 (6E-3)& 2E-3 (5E-3)& 1E-3 (2E-3)& 5E-4 (2E-3)& 3E-4 (1E-3)\\
ID-FW &3E-1 (9E-2)& 9E-2 (5E-2)& 3E-2 (2E-2)& 1E-2 (1E-2)& 5E-3 (9E-3)& 5E-3 (9E-3)& 3E-3 (6E-3)& 2E-3 (5E-3)& 1E-3 (2E-3)& 5E-4 (2E-3)& 3E-4 (1E-3)\\
HUB &3E-2 (1E-2)& 3E-2 (1E-2)& 4E-2 (9E-3)& 4E-2 (9E-3)& 4E-2 (8E-3)& 5E-2 (7E-3)& 4E-2 (7E-3)& 4E-2 (6E-3)& 4E-2 (6E-3)& 4E-2 (5E-3)& 4E-2 (4E-3)\\
KUR-ILP &3E-2 (1E-2)& 3E-2 (1E-2)& 4E-2 (9E-3)& 4E-2 (9E-3)& 4E-2 (8E-3)& 5E-2 (7E-3)& 4E-2 (7E-3)& 4E-2 (6E-3)& 4E-2 (6E-3)& 4E-2 (5E-3)& 4E-2 (4E-3)\\
COP-KM &2E-1 (6E-2)& 1E-1 (5E-2)& 8E-2 (3E-2)& 7E-2 (3E-2)& 6E-2 (2E-2)& 6E-2 (2E-2)& 5E-2 (1E-2)& 5E-2 (1E-2)&  &  &  \\
MPC-KM &3E-1 (7E-2)& 2E-1 (7E-2)& 1E-1 (4E-2)& 8E-2 (3E-2)& 8E-2 (2E-2)& 7E-2 (3E-2)& 7E-2 (2E-2)& 7E-2 (2E-2)&  &  &  \\
EM &5E-3 (9E-3)& 5E-3 (8E-3)& 3E-3 (6E-3)& 2E-3 (4E-3)& 8E-4 (2E-3)& 9E-4 (2E-3)& 5E-1 (1E-2)& 5E-1 (1E-2)& 5E-1 (7E-3)&  &  \\
LCVQE &3E-1 (7E-2)& 3E-1 (6E-2)& 2E-1 (6E-2)& 2E-1 (6E-2)& 2E-1 (5E-2)& 2E-1 (5E-2)& 2E-1 (6E-2)& 2E-1 (6E-2)& 2E-1 (5E-2)& 2E-1 (6E-2)& 2E-1 (5E-2)\\
CCLS &4E-2 (3E-2)& 7E-2 (3E-2)& 1E-1 (5E-2)& 3E-1 (6E-2)& 3E-1 (4E-2)& 3E-1 (3E-2)& 4E-1 (3E-2)& 4E-1 (3E-2)& 4E-1 (2E-2)&  &  \\
R100 &4E-1 (4E-2)& 4E-1 (3E-2)& 5E-1 (2E-2)& 5E-1 (2E-2)& 5E-1 (1E-2)& 5E-1 (1E-2)& 5E-1 (1E-2)& 5E-1 (1E-2)& 5E-1 (7E-3)& 5E-1 (5E-3)& 5E-1 (3E-3)\\
ID &5E-1 (6E-2)& 5E-1 (4E-2)& 5E-1 (2E-2)& 5E-1 (2E-2)& 5E-1 (2E-2)& 5E-1 (1E-2)& 5E-1 (1E-2)& 5E-1 (1E-2)& 5E-1 (7E-3)& 5E-1 (5E-3)& 5E-1 (3E-3)\\
   \hline
\end{tabular}
}
\caption{Simulations: optimization performance in the matching problem \eqref{MDADC}.
The relative error  (average across 100 replications) is displayed
 with standard deviation in parentheses. From  top to bottom of the table: best to worst performance. 
 Missing values are due to execution timeout (running time $>300s$).}
 \label{table:optimization}
\end{table}

 R100-BCA is the best method for each $n$, attaining the best objective value in virtually every replication. 
 For small values $n\in \{ 5,10\}$, ILP and KUR-IQP also achieve best performance. 
 The next best methods are LBL-BCA-2X, HUB-BCA-2X, LBL-BCA, and HUB-BCA, 
 with a relative error decreasing from order $10^{-3}$ for $n=5$ to order $10^{-4}$ or $10^{-5}$ for $n=100$. 
 Recall that the LBL initialization is an oracle of sorts 
 since data labels are typically not available in matching problems. 
The other combinations of methods of this paper yield slightly higher yet comparable relative error 
that goes roughly from order $10^{-2}$ for $n=5$ to the range $(10^{-4},10^{-6})$ for $n=100$. 
As can be expected, the ID and REC initializations yield slightly worse performance 
whereas R100 provides the best results. 
BCA is less sensitive to the initialization methods than FW and KM. 
EM, which is initialized with ID-BCA, gives reasonable results for $n\le 50$ (relative error of order $10^{-3}$) 
although it does not improve upon BCA. For $n > 50$ however its performance with respect to \eqref{MDADC}
severely deteriorates and its relative error climbs to about 0.4. 

Among the competitor methods, KUR-ILP has the best performance, 
with a relative error of order $10^{-2}$ across values of $n$. 
COP-KM and MPC-KM have relative errors 
that decrease from order $10^{-1}$ for small $n$  to order $10^{-2}$ for $n=100$. 
LCVQE has a slowly decreasing relative error that goes from 0.3 for $n=5$ to 0.2 for $n=100$. 
CCLS sees it relative error increase from order $10^{-2}$ for small $n$ to 0.4 for $n=100$.


\emph{Rand index.} 
The Rand index (RI) is a measure of agreement between two partitions of a set; it is suitable for matching problems 
which produce clusters and not individual label predictions. 
Here the data partition produced by a matching method is compared 
to the partition induced by the data classes, i.e. their underlying digits in $\{0,\ldots, 9\}$. 
While the goal of matching is to produce homogeneous data clusters 
and not to maximize agreement between the produced clusters 
and some underlying class/LBL-induced clusters, 
these two goals are aligned in the simulations because 
data vectors generated by a same digit class tend to be 
much closer to each other than to vectors generated by other digit classes.  
  
  
Given a set $D$ of  size $n$ and two partitions $X$ and $Y$ of $D$ into clusters, 
the RI  is defined as the ratio $(a+b)/ {n \choose 2}$, 
where $a$ is the number of pairs of elements in $D$ 
that are in a same cluster  both in $X$ and $Y$, 
and $b$ is the number of pairs of elements in $D$ 
that are in different clusters both in $X$ and $Y$. 
This can be interpreted as the fraction of correct decisions 
to assign two elements of $D$ either to the same cluster or to different ones.

\begin{figure}[ht]
\begin{center}
\includegraphics[scale=0.75]{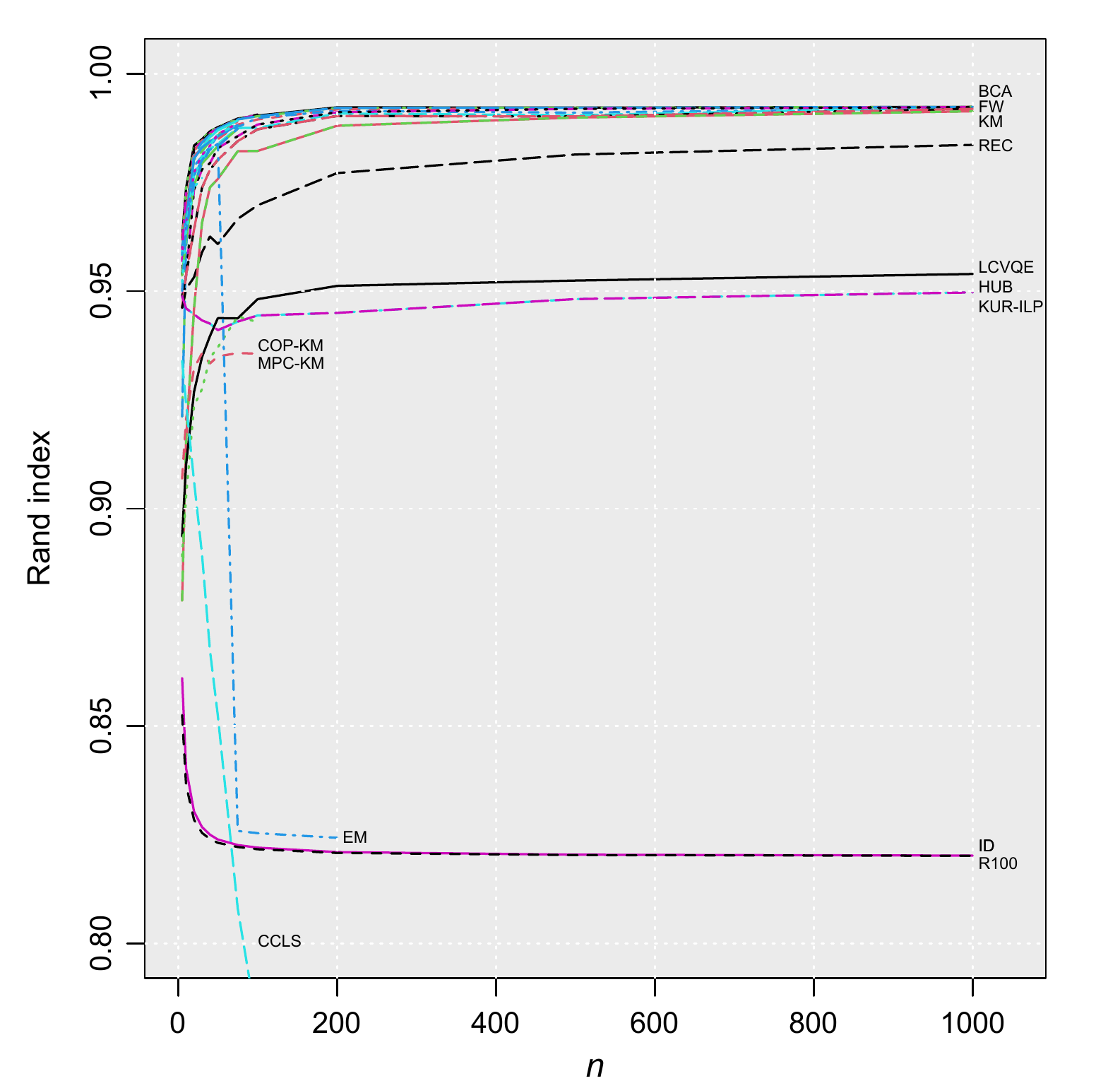}
\caption{Rand index versus sample size $n$ (average across 100 replications).}
\label{fig:rand}
\end{center}
\end{figure}

The RI of each method (averaged across 100 replications) 
is displayed in Figure \ref{fig:rand} as a function of $n$. Values closer to 1 
indicate better agreement between matching outputs and data labels (digits). 
For BCA, FW, and KM, the RI starts from a baseline in the range $[0.92,0.96]$, 
reaches 0.99 around $n=100$, and then stays at this level for $n > 100$.  
The REC initialization has a RI that increases from 0.94 for $n=5$ 
to 0.98 for $n=1000$. 
For COP-KM, MPC-KM, LCVQE, KUR-ILP, and HUB, 
the RI slowly increases from about 0.9 to 0.95 with $n$.
R100 and ID are two initializations nearly or full independent 
of the data labels, which are randomly shuffled. 
They are tantamount to random guessing 
and their baseline RI of  0.82 matches 
its theoretical expectation ($1- (2m-2)/m^2$).  
EM and CCLS show a RI that rapid decreases 
at or below random guessing levels, in accord with their 
modest performance in the optimization \eqref{MDADC}.

\begin{figure}[ht]
\begin{center}
\includegraphics[scale=0.75]{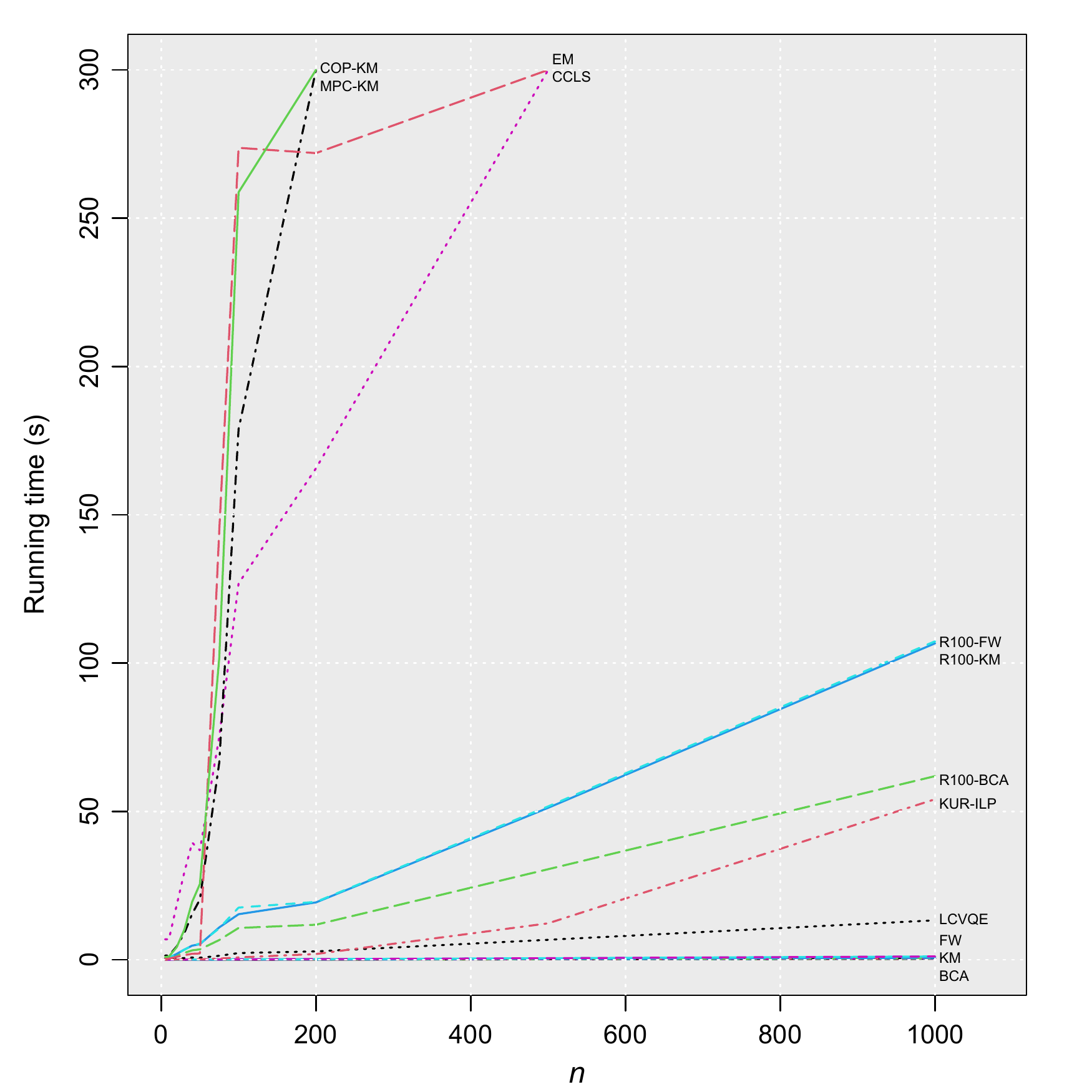}
\caption{Running time versus sample size $n$ (average across 100 replications).}
\label{fig:timing}
\end{center}
\end{figure}

\emph{Running time.} 
The running times of the algorithms are displayed in Figure \ref{fig:timing}. 
 During the simulations, algorithms were given 300 seconds to complete execution, after which they were interrupted.  
 Accordingly any value 300s on the figure (often largely) underestimates the actual computation time. 
 The algorithms can be divided in two groups: 
 those who can solve \eqref{MDADC} for $n=1000$ in 100s or far less, 
 and those that time out (execution time over 300s) for $n\le 500$ or far less.  
 They are described below by order of decreasing speed.

BCA, FW, and KM are the fastest methods with running times 
of order $10^{-3}$ to $10^{-1}$ seconds across values of $n$. 
For $n=1000$, they are one order of magnitude faster than the next best method (LCVQE). 
The HUB and REC initializations, 
although slower than arbitrary starting points like identity or random permutations, 
enable overall faster computations because good starting points 
reduce the number of iterations required for the main algorithm to converge.
Completion of (100 runs) of BCA, FW, or FW based on the R100 initialization 
takes between 200 and 250 times the execution of a single run based on HUB or REC (instead of roughly 100). 
This is because the latter heuristics find good starting points  
whereas some (or many) of the 100 random starting points
 will be bad and require many iterations for the main algorithm to converge. 
KUR-ILP enjoys the same speed as the BCA, FW, and KM 
for small $n$ but its running time appears to scale polynomially with $n$.  
LCVQE appears to scale linearly with $n$ but with a much larger multiplicative constant than 
BCA, FW, and KM. Its running time is of order $10^{-2}$s for $n=5$ and 1s for $n=75$. 
The running time of CCLS grows steeply with $n$ and exceeds the 300s limit for $n\ge 500$. 
MPC-KM, COP-KM and EM are very slow, at least in their R implementation, 
and they time out (i.e. their execution times exceed 300s) for $n\ge 200$. 
Their computational load seems to grow exponentially with $n$. 
In the case of the EM, the computational bottleneck 
is the evaluation of matrix permanents.
ILP, KUR-IQP and 2X are by far the slowest methods in the simulations. 
The first two stall and time out as soon as $n$ exceeds a few units, 
although they produce good results when $n\le 5$. 
The computational load of 2X scales exponentially with $n$ 
(average computation time 110s for $n=30$); 
it is much higher when using the ID, R100, REC, and HUB 
initializations than when applied as a post-processing step 
following, say, the BCA method.

\emph{Summary of simulations.} 
\begin{itemize}
\setlength\itemsep{-1mm}
\item[-] BCA is the fastest and most accurate of all studied methods. 
It provides excellent accuracy when initialized with REC or HUB.  
For best accuracy, the R100 initialization should be used at the cost of 
increased yet still manageable  computations. 

\item[-] BCA, KM, and FW are overall extremely fast and can handle  
datasets of size $n = 10^3$ and up without difficulty. 
KM and FW are slightly less accurate than BCA in terms of optimization performance 
(relative error between $10^{-3}$ and $10^{-4}$) and Rand index.  

\item[-] 2X is computationally costly and fairly inaccurate when used on its own, i.e. with an arbitrary starting point. 
It largely improves the accuracy of KM and FW solutions but not of BCA solutions. 
It is mostly beneficial in small to moderate dimension $n$.  

\item[-] HUB and REC are not sufficiently accurate to be used on their own 
but they provide good starting points to more sophisticated matching methods. 
HUB uses data more extensively than REC and yields slightly better performance. 

\item[-] For moderate to large $n$, EM shows poor performance in both computations 
(due to the evaluations of matrix permanents) and optimization. 
Its performance is satisfactory for $n\le 50$, possibly because of the BCA initialization. 

\item[-] ILP and KUR-BQP are only computationally feasible in very small samples ($n\le 10$ or so). 
In this setup they virtually always find the global minimum of \eqref{MDADC}. 

\item[-] KUR-ILP is relatively fast (it solves \eqref{MDADC} for $n=1000$ in 50s) 
but not highly accurate (relative error between  3\% and 5\%).   
LCVQE  is both faster and far less accurate:  
 it solves \eqref{MDADC} for $n=1000$ in 13s but has relative error 
 in $(0.2,0.3)$ for all values of $n$. 
 
\item[-] COP-KM and MPC-KM have very similar profiles in computation time and optimization accuracy. 
Their relative error goes from 0.2-0.3 for $n=5$ to 0.05-0.06 for $n=100$. 
They are not able to handle large datasets (at least not in their R implementation) 
as their computations stall for $n\ge 200$. 
CCLS only performs reasonably well for $n\le 10$. Its Rand index and relative error deteriorate 
quickly as $n$ increases and its computations time out for $n\ge 500$.

\end{itemize}


\section{Application to fMRI data} 
\label{sec:fmri}

In this section we harness the matching problem \eqref{MDADC} and its proposed solutions 
to analyze resting-state functional magnetic resonance imaging (rs-fMRI) data, 
the goal being to explore the dynamic functional connectivity (DFC) of the brain. 
In short, functional connectivity (FC) relates to the integration of brain activity, 
that is, how distant brain regions coordinate their activity to function as a whole. 
The dynamic nature of FC, in particular its dependence on factors such as task-related activity, psychological state, and cognitive processes, is well established in neuroimaging research \citep[e.g.][]{Chang2010,Handwerker2012,Hutchison2013}.  

The present analysis aims to extract measures of DFC from individual subject data 
and match these measures across subjects to uncover common patterns and salient features.  
The data under consideration are part of the ABIDE preprocessed data \citep{Craddock2013}, a large corpus of rs-fMRI measurements recorded from subjects diagnosed with autism spectrum disorder and from control subjects. These data and detailed descriptions are available at \href{http://preprocessed-connectomes-project.org/abide/}{preprocessed-connectomes-project.org/abide/}. 
We selected the following preprocessing options: Connectome Computation System (CCS) pipeline,  spatial averaging over 116 regions of interest (ROI) defined by the AAL brain atlas, bandpass temporal filtering, no global signal regression. For simplicity, we only used data from control subjects and discarded data that did not not pass all quality control tests. 
 This resulted in $n=308$ subjects with fMRI time series of average length about 200 scans (SD=62).

 \paragraph{Subject-level analysis.} 
Vector autoregressive (VAR) models are widely used to assess FC in fMRI data
\citep{Valdes-Sosa2005,Friston2013,Ting2017}. 
Here we represent
 the fMRI time series of a subject 
  by a piecewise VAR model of order 1: 
\begin{equation}\label{VAR1}
y_t = A_t y_{t-1} +b_t +  \varepsilon_t  \qquad (1 \le t \le T)
\end{equation}
where  $y_t $ is an fMRI measurement vector of length 116,
$A_t$ an unknown regression matrix encoding FC dynamics, 
$b_t$  an unknown baseline vector, and $\varepsilon_t $ 
 a random noise vector with multivariate normal distribution $N(0,Q_t)$.  
The $A_t$ are assumed sparse, 
reflecting the fact that only a small number of ROIs at time $t-1$ are predictive of ROI activity at time $t$. 
The model parameters  $( A_{t},b_t, Q_t )$ are assumed piecewise constant with few change points, 
indicating that FC states persist for some time (say, between 5 and 50 scans) before the brain switches to a different FC state. 

For each subject, the task at hand is to simultaneously detect change points 
in \eqref{VAR1} and estimate $(A_{t}, b_t)$ over the associated time segments. 
($Q_t$ is of secondary importance here and can be ignored). 
The sparse group fused lasso (SGFL) approach of \cite{Degras2020} is designed for this purpose. 
To simplify the task of determining a suitable range for the SGFL regularization parameters 
and calculating regularization paths, we employ the two-step procedure of this paper. 
The first step detects change points via the group fused lasso \citep[e.g.][]{Bleakley2011};  
the second step recovers sparse estimates of the $A_t$ 
separately on each segment 
via the standard lasso \citep{Tibshirani1996}. 

After fitting the regularization paths, a single lasso estimate  $(\hat{A}_t,\hat{b}_t)$ 
is selected for each segment by the Akaike Information Criterion.   
Among all generated model segmentations, we retain 
the one with the most segments satisfying the following criteria: 
(i) \emph{length}: the segment must have at least 5 scans, 
(ii) \emph{goodness of fit}: the lasso fit must have a deviance ratio at least 0.3, 
and (iii) \emph{distinctness}: the parameter estimate $\hat{A}_t$ for the segment must have at least 10\% relative difference 
with estimates of other selected segments. 
To facilitate interpretation and remove noisy components,  10 segments are retained per subject at the most.

\paragraph{Group-level analysis.}
 Following the subject-level analysis, a set of change points and associated model parameter estimates 
 is available for each subject, 
say $\{(\hat{A}_{ik}, \hat{b}_{ik} , \hat{T}_{ik} ): 1\le k \le m_i \}$ 
with $\hat{T}_{ik}$  the $k$th change point 
and $m_i$ the number of segments
for the $i$th subject ($1\le i\le n$).  
The regression matrices $\hat{A}_{ik}$ provide informative  FC measures 
and could  in principle be used for group-level comparisons. 
They are however highly sparse 
and matching them using the squared Euclidean distance of problems 
\eqref{MDADC}-\eqref{objective unbalanced} does not produce sensible results.  
We thus calculate the empirical correlation matrices on each segment
$\{  \hat{T}_{ik} , \ldots,  \hat{T}_{i(k+1)} -1 \}$ and take them as inputs 
for the group analysis. 
See e.g. \citep{Wang2014} for a review of common FC measures in neuroimaging. 
 After discarding correlation matrices based on short segments 
 (10 scans or less, for increased estimation accuracy) and 
extracting the lower halves of the remaining matrices,  
we obtain a set $\{ x_{ik} : 1\le i \le 306, 1\le k \le m_i \}$
of 1801  correlation vectors of size $p= 116 \times 115 / 2 = 6670$.  
The number $m_i$ of vectors per subject varies in the range $[ 1,10]$  
with an average of 5.88 (SD=1.77). 
The unbalanced matching problem \eqref{objective unbalanced} 
is then solved for $K\in\{ 10, 20, \ldots, 100\}$  
using a generalized version of the BCA Algorithm \ref{alg:bca}.
Based on the inspection of the cluster centers and cluster sizes, 
we retain the matching based on $K=100$ clusters. 
With this choice, cluster sizes are in the range $[12,28]$ (mean=18.01, SD=4.16). 
Smaller values of $K$, say $K\ge 50$, would be equally fine for data exploration. 
$K$ should however not be too small so as to avoid large clusters 
in which fine details of FC are averaged out in the cluster center and only large-scale features remain.

\begin{figure}[ht!] 
\begin{center}
\includegraphics[width= \textwidth]{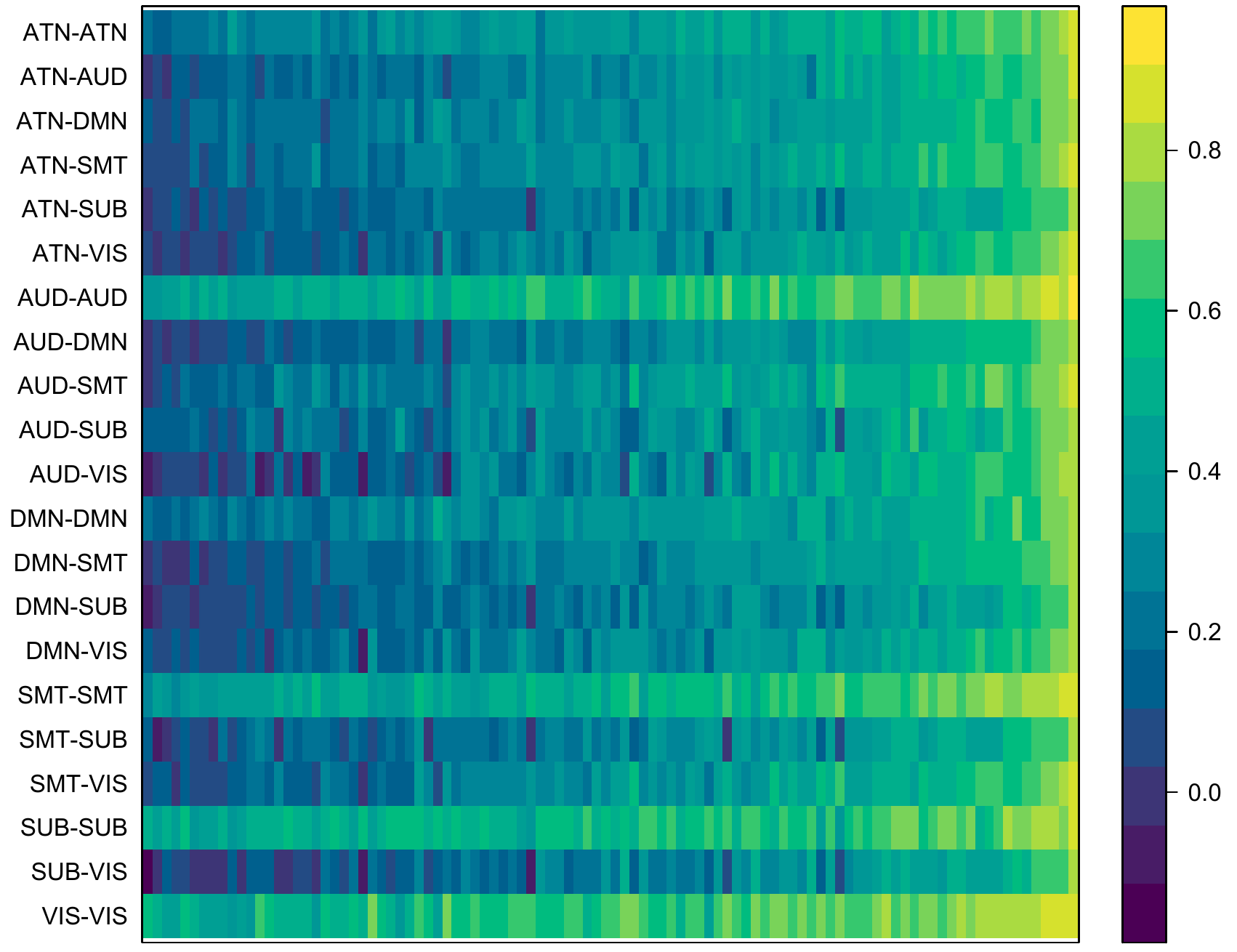} 
\end{center}
\caption{rs-fMRI data analysis. 
Each column represents the center of a cluster of matched features, that is, 
(half) a correlation matrix averaged across cluster members (subjects) 
and across ROIs of resting state networks. ATN: attentional network, AUD: auditory network, 
DMN: default mode network, SMT: sensorimotor network, SUB: subcortical network, VIS: visual network.} 
\label{fig:fmri all centers}
\end{figure}

Figure \ref{fig:fmri all centers} displays the 100 resulting cluster centers, 
i.e. the average correlation matrices of the clusters. 
For easier visualization and interpretation, the ROI-level correlations are aggregated 
into six well established \emph{resting state networks} (RSN): 
 the  attentional network (26 ROIs), auditory network (6 ROIs), 
 default mode network (32 ROIs), sensorimotor network (12 ROIs), subcortical network (8 ROIs), 
 and visual network (14 ROIs). A list of the ROI names and associated RSNs is given in Appendix \ref{sec:appendix}. 
 Note that some ROIs do not belong to any known functional networks while others are recruited in two networks. 
 %
The visual network and auditory network have strong intracorrelation (0.59 and 0.64 on average across cluster centers, respectively, not including TPOsup in the auditory network). 
The subcortical network and sensorimotor network show moderate internal correlation (0.51 on average each). 
The default mode and attentional networks comprise more ROIs and are usually less correlated (0.36 and 0.40 on average, respectively). 
The hippocampus (HIP), parahippocampal gyrus (PHG), and amygdala (AMYG) cluster together fairly strongly (average correlation 0.53). Applying community detection algorithms to each cluster center with the R package \verb@igraph@, we noticed that 
ROIs from the visual network are virtually always in the same community; the same holds true for the subcortical network.
The strongest correlations found between RSNs are the following: auditory--sensorimotor (0.38 on average across clusters) attentional--default mode (0.36), attentional--sensorimotor (0.36), and sensorimotor--visual (0.35).

A remarkable feature (not visible in Figure \ref{fig:fmri all centers}) is the strong positive correlation 
between the Rolandic Operculum (ROL) and the regions PUT (subcortical network), PoCG, PreCG, and SMG (sensorimotor), and HES, STG (auditory) (between 0.42 and 0.67). In addition, CB9.R, VERMIS10, CB10.R, PCG.L, VERMIS9 exhibit consistent negative correlation (or at least lower average correlation) with most other ROIs. In particular, CB9.R (cerebellum) has 36.5\% of negative correlations with other ROIs  whereas the overall proportion of negative correlations in the 100 average correlation matrices is only 10.6\%.

\begin{figure}[ht!] 
\vspace*{-5mm}
\begin{center}
\includegraphics[width=.45 \textwidth]{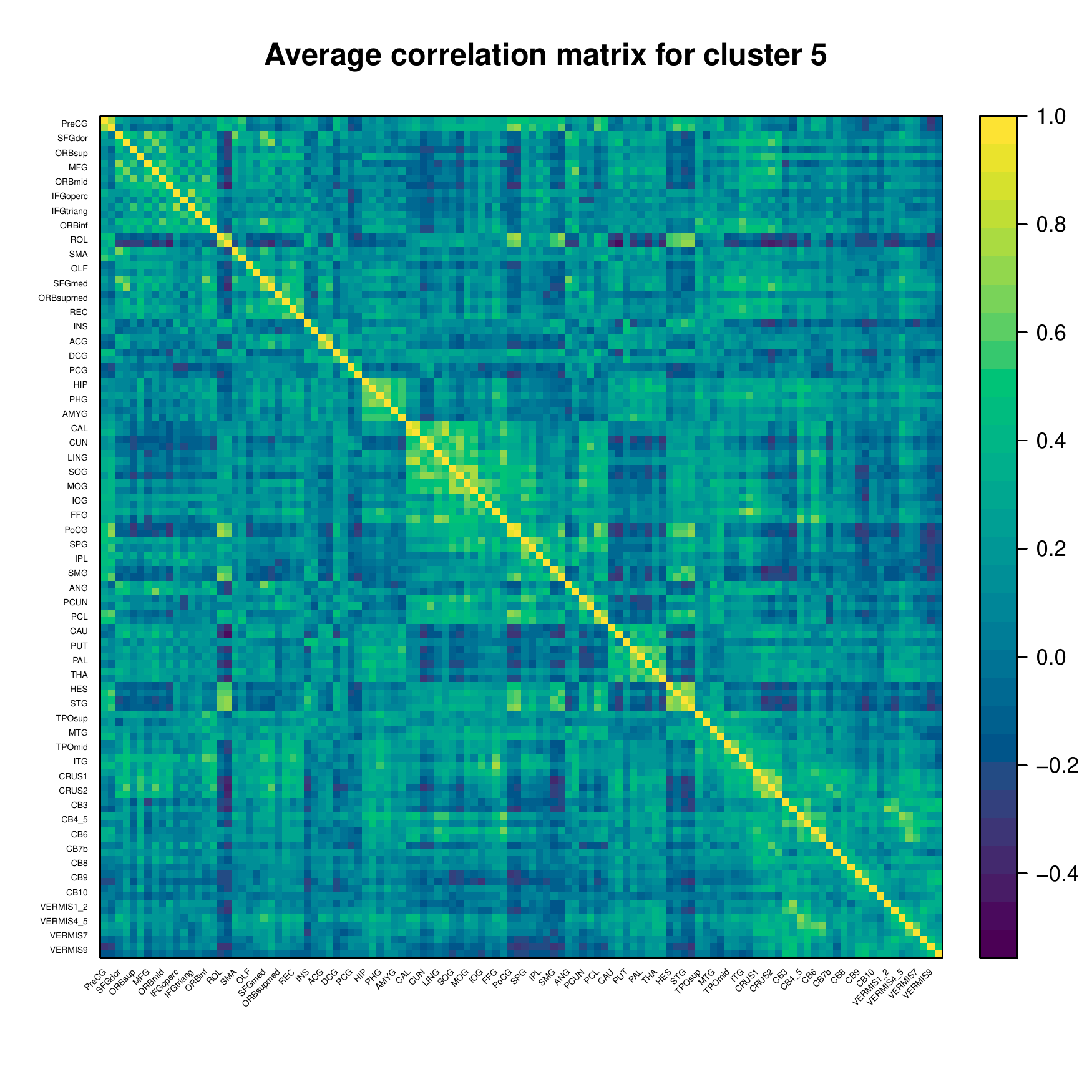} 
\includegraphics[width=.45 \textwidth]{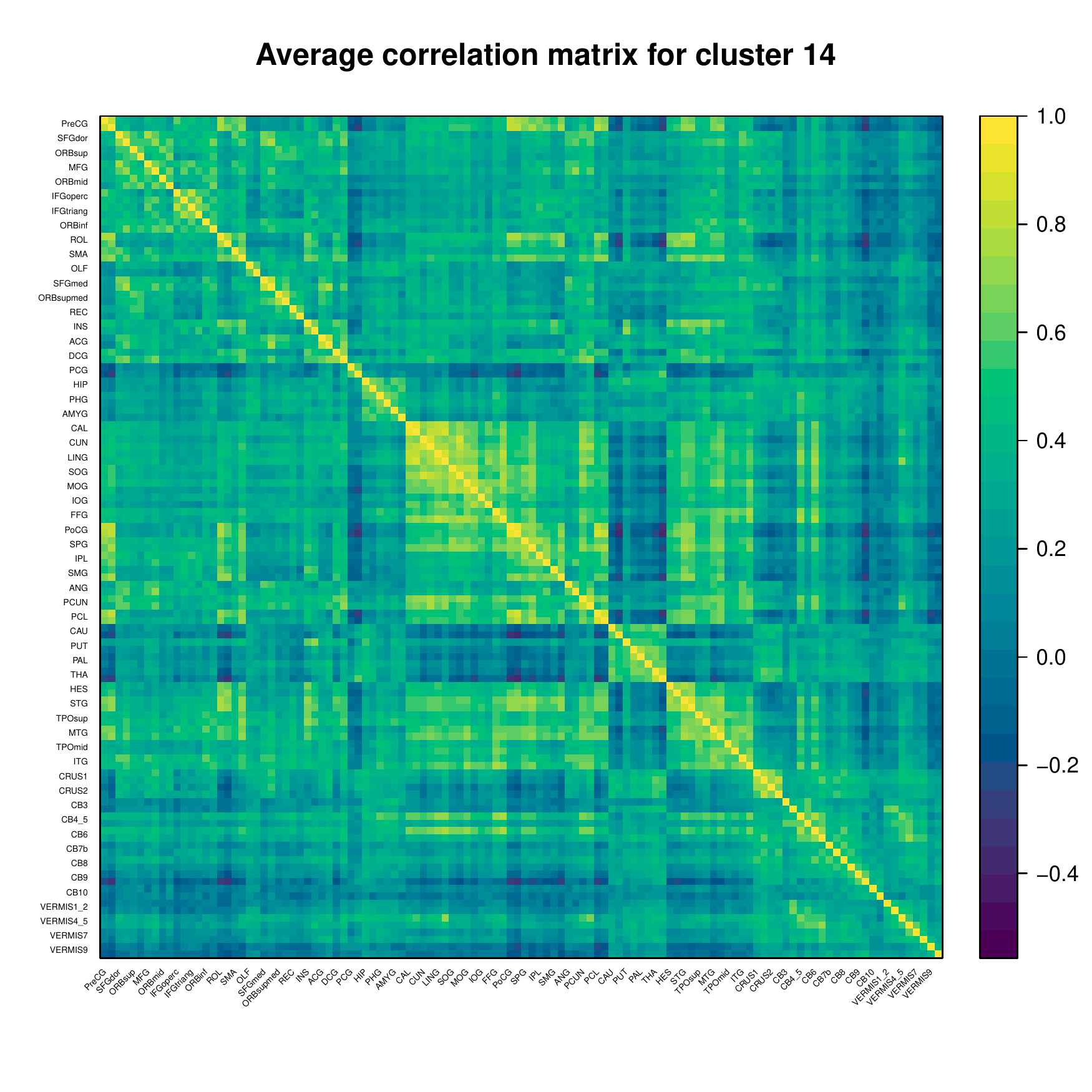} \\
\vspace*{-6mm}
\includegraphics[width=.45 \textwidth]{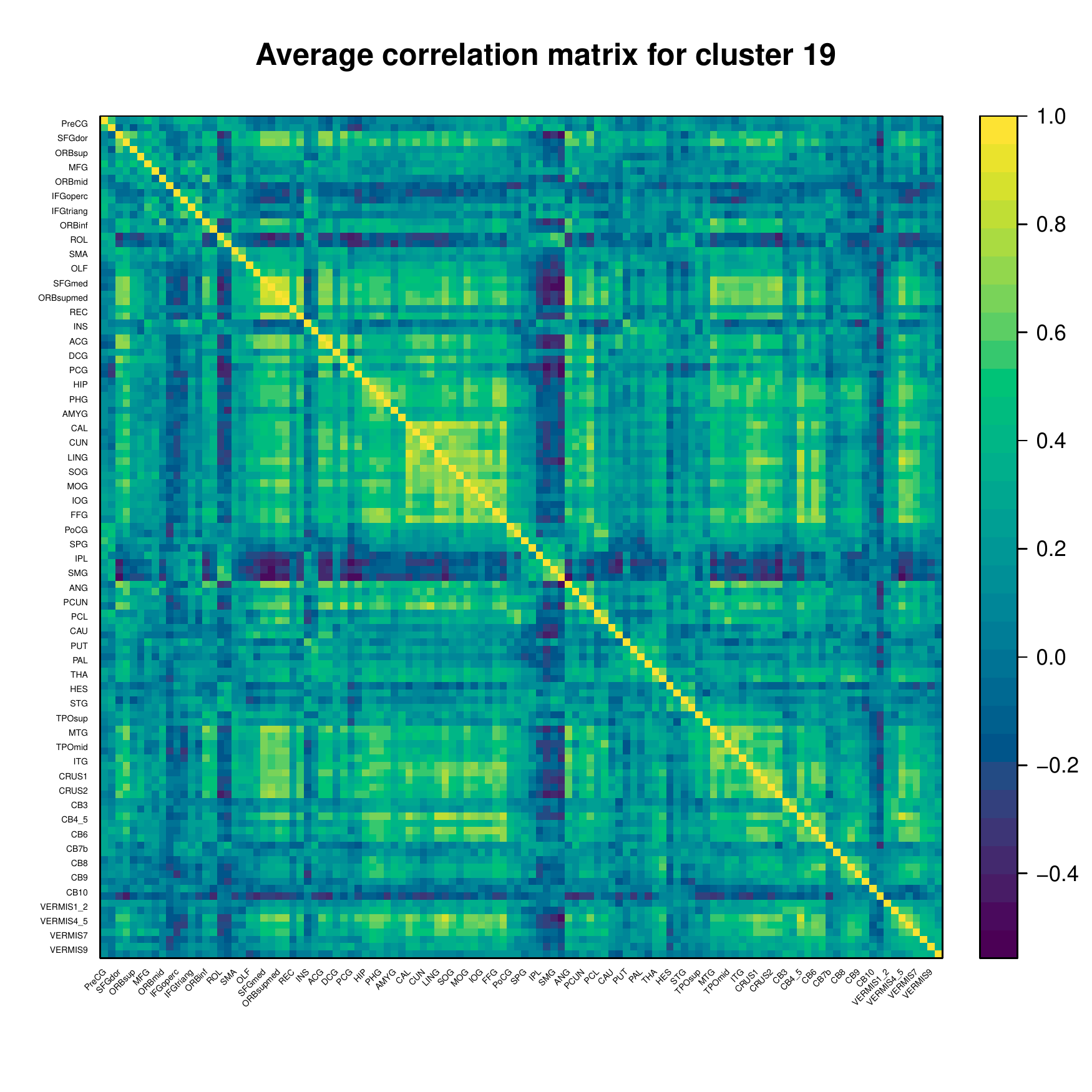} 
\includegraphics[width=.45 \textwidth]{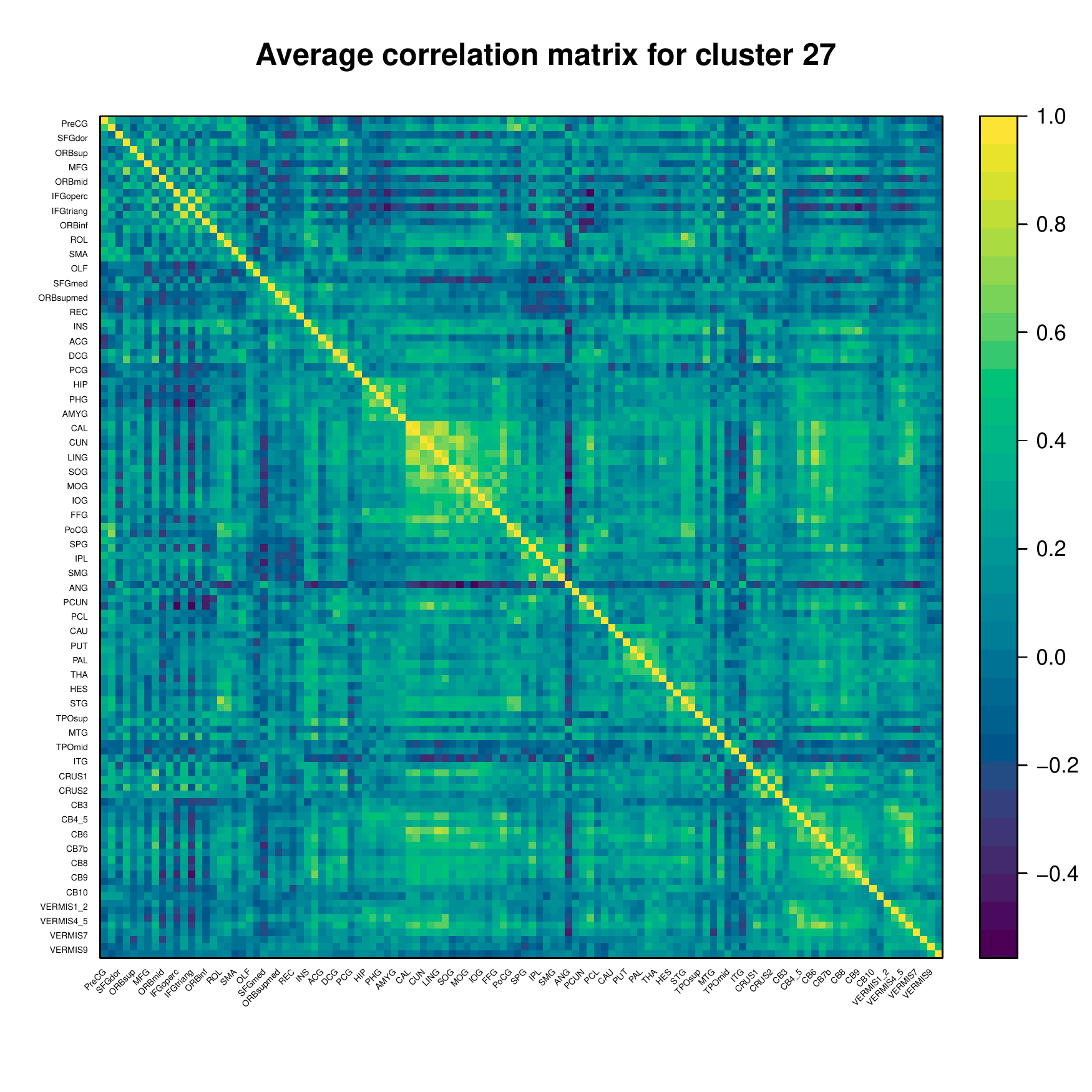} \\
\vspace*{-6mm}
\includegraphics[width=.45 \textwidth]{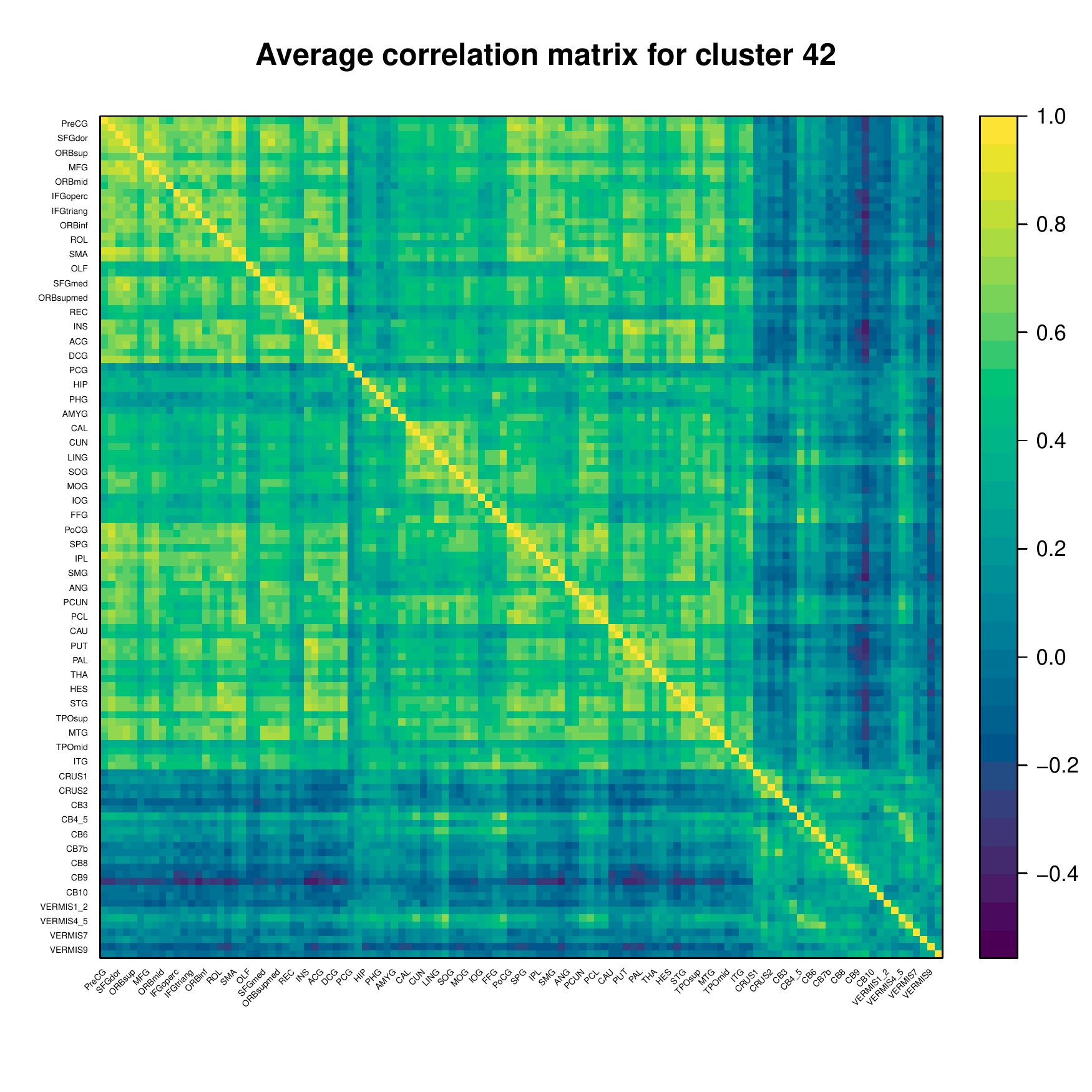} 
\includegraphics[width=.45 \textwidth]{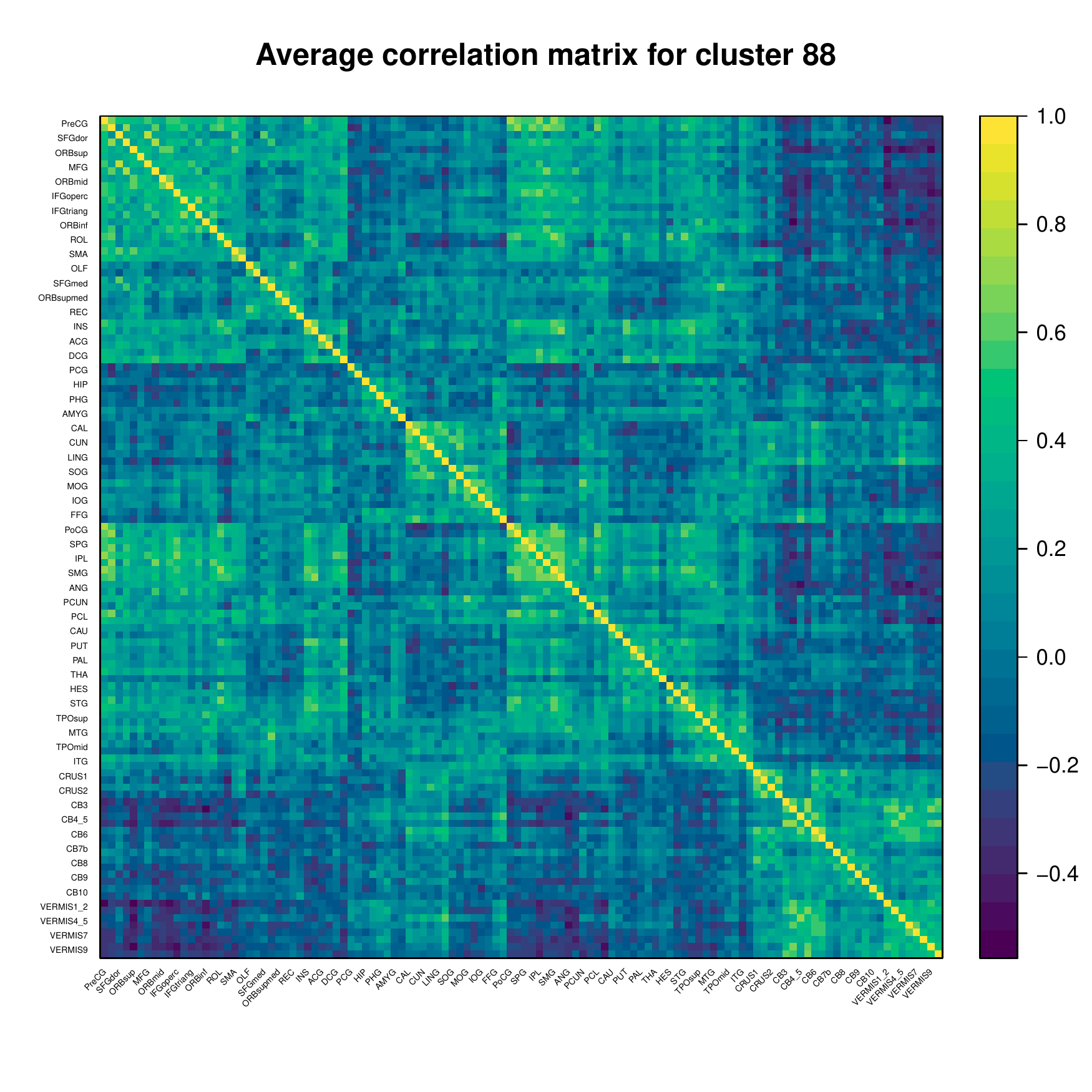} 
\end{center}
\vspace*{-7mm}
\caption{rs-fMRI data analysis. Examples of cluster centers (averages correlation matrices) 
derived from matching individual correlation matrices  across subjects.  
Each displayed matrix corresponds to a cluster of 14 to 23 subjects.} 
\label{fig:fmri dfc}
\end{figure}

Figure \ref{fig:fmri dfc} shows interesting examples of average correlation matrices (cluster centers) at the ROI level. 
 Cluster 5 shows strong positive correlation within the 
auditory, 
subcortical,  
and visual networks, 
and in the small groups (HIP, PHG, AMYG),  
(CRUS1, CRUS2), and (CB3--CB6, VERMIS1--VERMIS7). 
ROL has moderate to strong negative correlation with CRUS1, CRUS2
and regions from the subcortical network (dark blue stripe towards the top and left) 
and  strong positive correlation with PoCG, SMG (sensorimotor) and HES, STG (auditory). 
The auditory and sensorimotor networks have moderate to strong positive correlation. 
 Cluster 14 shows clear blocking structure along the diagonal (correlation within RSN) as well as 
 anticorrelation patterns between 
CAU, PUT, PAL, THA (subcortical) and  ROL, PoCG (sensorimotor), PCL (sensorimotor);  
and between PCG (default mode) and PreCG (sensorimotor), ROL, PoCG (sensorimotor), PCL (sensorimotor). 
Community detection reveals three large and heterogeneous communities (sizes 43, 40, 36). 
Cluster 19 displays moderate to strong negative correlation (-0.55,-0.25) between  
IPL, SMG, ROL, CB10.R on the one hand and about 40 other ROIs on the other. 
The alternating clear and dark lines in cluster 27 reveal lateralized anticorrelation patterns between ROIs in the attentional network on the left side of the brain with most other ROIs in the brain. 
Cluster 42 shows  two roughly uncorrelated blocks, 
a very large one with strong intracorrelation 
and a  smaller one (CRUS, CB, VERMIS) with weaker intracorrelation. 
Cluster 88 displays a checkered correlation structure with 
strong anticorrelation between (CRUS, CB, VERMIS) and the rest of the brain.

\emph{Summary of the data analysis.} 
The data analysis has established that the matching approach \eqref{MDADC}-\eqref{objective unbalanced} 
provides scientifically meaningful insights into DFC at the group level. 
By inspecting the cluster centers (average correlation matrices) produced by the matching process, 
one recovers large-scale patterns consistent with neuroscientific knowledge. 
For example, known resting state networks are clearly reflected 
in the blocking structure of the cluster centers (see Figure \ref{fig:fmri dfc}). 
But the cluster centers can also generate new insights and hypotheses.  
For example, the Heschl gyrus (HES) is not systematically included 
in the auditory network but, according to our analysis, it should. 
Similarly, the ROI TPOsup (temporal lobe: superior temporal gyrus), 
although it is near to or part of the auditory cortex, 
has shown only weak correlation with the other ROI of the auditory network,  
Superior temporal gyrus (STG). These elements may lead to a more nuanced 
understanding of the auditory network.  
Other remarkable findings include the strong anticorrelations found between 
the Rolandic operculum (ROL), the cerebellum (CER) and the vermis (VERMIS) 
on the one hand and (a large part of) the rest of the brain on the other. 
Importantly, by design, 
each of the clusters formed by the matching process highlights 
commonalities \emph{between} subjects and not \emph{within} subjects. 
This is in contrast with unconstrained clustering methods (e.g. $K$-means clustering) 
whose clusters may consist in (vectors from) a small
number of or even a single subject in extreme cases. 
%


\section{Discussion} 
\label{sec:discussion}

We have sought to efficiently match feature vectors 
in a one-to-one fashion across large collections of datasets or statistical units.   
 In  applications where statistical units are matched in pairs, 
this task is conveniently framed as a multidimensional assignment problem with decomposable costs (MDAC).  
Taking the squared Euclidean distance as  dissimilarity function in the MDADC  
enables tremendous computational speedup by transforming ${n\choose 2 }$ related matching problems between all pairs of datasets into  $n$ separate matching problems between each dataset and a template. 
Leveraging this idea, we have developed extremely fast algorithms 
whose computational complexity scales linearly with $n$.  
These algorithms do not require precalculating and storing 
assignment costs, which may be infeasible in large-scale applications. 
Instead, they efficiently calculate assignment costs on the fly.  
To our knowledge, no other available method to solve the MDADC possesses 
either of these linear scaling and storage-free properties necessary to large-scale applications.  

Our proposed algorithms rely on various optimization techniques such as $K$-means clustering, block coordinate ascent (BCA), 
convex relaxation, the Frank-Wolfe algorithm, and pairwise interchange heuristic.   
We have also taken a probabilistic view of \eqref{MDADC} leading to a constrained Gaussian mixture model 
and associated EM algorithm. 
Altogether the proposed algorithms form a panel that covers most types of approach to the MDADC found in the literature. 
(As discussed earlier, we have not considered Lagrangian relaxation methods 
as they require computer clusters and/or GPUs for efficient large-scale implementation.)  
These algorithms extend or specialize existing approaches in a nontrivial way. 
For example, the BCA and 2-exchange algorithms, which are specialized versions of existing algorithms, 
 scale linearly with $n$ and are amenable to large-scale applications whereas the more general algorithms are not.

The numerical study 
has shown the excellent performances of the three main algorithms: $K$-means matching, BCA, and Frank-Wolfe, with respect to computation and optimization. In particular, these algorithms largely outperform all competitors  and can handle  very large collections of data. The BCA algorithm shows slightly better performance than the other two. The pairwise interchange heuristic can enhance these two methods to reach near optimality at a hefty computational price. 
The EM algorithm displayed fairly poor performance throughout the study. Upon inspection, the poor optimization results  came from the fact that the algorithm was ``too sure" about the allocation probabilities (of data vectors to classes) which were almost invariably calculated as 0 or 1. This in turn may arise from the (relatively) high dimension of the data, short tails of the normal distribution, and/or error in  covariance estimation. Using the deterministic annealing EM and/or random starting points did not fix the issue. Solutions for improving the EM optimization may be to impose a diagonal structure on covariance estimates or to consider (mixtures of) distributions with heavier tails such as 
multivariate $t$-distributions. The computational slowness of the EM could be remedied by calculating a small fixed number of most likely allocations rather than computing them all through matrix permanents.  


The analysis of the ABIDE preprocessed fMRI data 
has shown the strong potential of the proposed feature 
matching approach for exploring neuroimaging biomarkers
and producing interpretable clusters at the group level. 
A key characteristic of one-to-one feature matching 
is that, unlike unsupervised clustering, 
it is guaranteed to produce ``representative" clusters 
that reflect variations between subjects and not within. 
While feature matching was employed in our analysis
 for data exploration, 
this technique could also be used in a more principled way 
as a preliminary step to disentangle association ambiguities between 
biomarkers and/or to stratify subjects into small, homogenous groups 
prior to a group-level analysis. Such matching-based approach could be for example 
compared to the consensus clustering strategy of  \cite{Rasero2019}.

\paragraph{Possible extensions and future work.} 

\begin{itemize}
\item \emph{Weighted (squared) Euclidean distance.} 
The squared Euclidean distance in \eqref{MDADC} can easily be generalized to a weighted squared Euclidean distance 
$\| x \|_W^2 = x ' W x$ with $W\in \mathbb{R}^{p\times p}$ a positive semi-definite matrix. Decomposing $W$ as $L'L$ (e.g. by Cholesky decomposition), it suffices to premultiply each matrix $X_i$ by $L$ to formulate an equivalent problem \eqref{MDADC} 
using the unweighted (squared) Euclidean distance. 
  
\item \emph{Alternative dissimilarity measures.} 
Although the squared Euclidean distance for $d$ in the general MDADC problem \eqref{MAP}-\eqref{DC} enables extremely fast and scalable algorithms with low memory footprint, 
it may not adequately capture relevant differences between feature vectors in some applications.  
If the Euclidean distance $\| \cdot \|_2$
or the Manhattan distance $\| \cdot \|_1$, for example, is a more sensible choice for $d$, a reasonable approach would be to use an objective function based on the ($nm$) distances between feature vectors and their cluster centers instead of one based on the distances between all ${ n\choose 2 } m$ pairs of matched vectors. 
In this case, the $K$-means matching Algorithm \ref{alg:kmeanslike} can be adapted as follows. 
The assignment step remains the same: given cluster centers $c_1,\ldots,c_m$, 
the feature vectors of each unit $i\in[n]$ are assigned to clusters by minimizing the LAP 
with assignment matrix  $A_{i} = ( d( x_{ik} , c_{l} ))_{1 \le k,l \le m}$. 
The updating step for the cluster centers proceeds from calculating $m$ geometric medians if $d=\| \cdot \|_2$, 
or $mp$ univariate medians id $d = \| \cdot \|_1$. Both these tasks can be accomplished in near linear time, 
and like in the case $d=\| \cdot \|_2^2$, no distance needs to be pre-calculated and stored. 
Accordingly, the modified objective function and modified $K$-means matching algorithm 
still enable linear time complexity linear in $n$ and low space requirements.
(The other algorithms of Section \ref{sec:methods} do not extend quite so nicely as they fundamentally 
rely on the scalar product and separability properties that underlie $\| \cdot \|_2^2$.)
%
%
%

%

\item Gaining theoretical understanding of the optimization properties of the algorithms of this paper, for example by establishing deterministic or probabilistic bounds on their performances, could maybe explain the very good performances observed 
and/or give insights on worst case performance in difficult instances \citep[e.g.][]{Gutin2008}. Also, obtaining tight lower bounds through suitable Lagrangian relaxations would be desirable in practice.

\item The rich structure of problem \eqref{MDADC} may make it possible to easily construct 
instances in which the global minimum and optimal assignment are known \cite[see e.g.][for related work on quadratic assignment problems]{Drugan2015}. This would  be of course useful to benchmark methods.

\end{itemize}

\subsection*{Acknowledgments}
The author  thanks to Vince Lyzinski for early discussions 
that led to the convex relaxation/\\Frank-Wolfe approach of Section \ref{sec:FW}. 
He also acknowledges his student Yiming Chen for his assistance in the literature search and the numerical study.


\bibliographystyle{plainnat}
\bibliography{matchFeat}

\clearpage
\appendix

\section{Brain regions of interest in fMRI data analysis}
\label{sec:appendix}


\begin{table}[H]
\vspace*{-0mm}
\begin{center}
\scalebox{.53}{
\begin{tabular}{clc    clc}
  \hline
  \hline
Label & Name & Abbrv & Label & Name & Abbrv \\ 
  \hline \hline
&  \multicolumn{2}{ l }{\textbf{Subcortical network}} &  & \multicolumn{2}{l}{\textbf{Default mode network}} \\ 
  71 & L Caudate nucleus & CAU.L & 5 & L Superior frontal gyrus, orbital & ORBsup.L \\ 
  72 & R Caudate nucleus & CAU.R & 6 & R Superior frontal gyrus, orbital & ORBsup.R \\ 
  73 & L Putamen & PUT.L & 7 & L Middle frontal gyrus & MFG.L \\ 
  74 & R Putamen & PUT.R & 8 & R Middle frontal gyrus & MFG.R \\ 
  75 & L Pallidum & PAL.L & 15 & L Inferior frontal gyrus, orbital & ORBinf.L \\ 
  76 & R Pallidum & PAL.R & 16 & R Inferior frontal gyrus, orbital & ORBinf.R \\ 
  77 & L Thalamus & THA.L & 23 & L Superior frontal gyrus, medial & SFGmed.L \\ 
  78 & R Thalamus & THA.R & 24 & R Superior frontal gyrus, medial & SFGmed.R \\ 
  \cline{1-3}
&  \multicolumn{2}{l}{\textbf{Auditory network}} 
   & 25 & L Superior frontal gyrus, medial orbital & ORBsupmed.L \\ 
  79 & L Heschl gyrus & HES.L & 26 & R Superior frontal gyrus, medial orbital & ORBsupmed.R \\ 
  80 & R Heschl gyrus & HES.R & 31 & L Cingulate gyrus, anterior part & ACG.L \\ 
  81 & L Superior temporal gyrus & STG.L & 32 & R Cingulate gyrus, anterior part & ACG.R \\ 
  82 & R Superior temporal gyrus & STG.R & 33 & L Cingulate gyrus, mid part & DCG.L \\ 
  83 & L Temporal pole: superior temporal gyrus & TPOsup.L & 34 & R Cingulate gyrus, mid part & DCG.R \\ 
  84 & R Temporal pole: superior temporal gyrus & TPOsup.R & 35 & L Cingulate gyurs, posterior part & PCG.L \\ 
     \cline{1-3}
   & \multicolumn{2}{l}{\textbf{Sensorimotor network}}  & 36 & R Cingulate gyrus, posterior part & PCG.R \\ 
  1 & L Precentral gyrus & PreCG.L & 37 & L Hippocampus & HIP.L \\ 
  2 & R Precentral gyrus & PreCG.R & 38 & R Hippocampus & HIP.R \\ 
  19 & L Supplementary motor area & SMA.L & 39 & L Parahippocampus & PHG.L \\ 
  20 & R Supplementary motor area & SMA.R & 40 & R Parahippocampus & PHG.R \\ 
  57 & L Postcentral gyrus & PoCG.L & 61 & L Inferior parietal gyrus & IPL.L \\ 
  58 & R Postcentral gyrus & PoCG.R & 62 & R Inferior parietal gyrus & IPL.R \\ 
  59 & L Superior parietal gyrus & SPG.L & 65 & L Angular gyrus & ANG.L \\ 
  60 & R Superior parietal gyrus & SPG.R & 66 & R Angular gyrus & ANG.R \\ 
  63 & L Supramarginal gyrus & SMG.L & 67 & L Precuneus & PCUN.L \\ 
  64 & R Supramarginal gyrus & SMG.R & 68 & R Precuneus & PCUN.R \\ 
  69 & L Paracentral lobule & PCL.L & 85 & L Middle temporal gyrus & MTG.L \\ 
  70 & R Paracentral lobule & PCL.R & 86 & R Middle temporal gyrus & MTG.R \\ 
  \cline{1-3}
&  \multicolumn{2}{l}{\textbf{Visual network}}  & 87 & L Temporal pole: middle temporal gyrus & TPOmid.L \\ 
  43 & L Calcarine fissure and surrounding cortex & CAL.L & 88 & R Temporal pole: middle temporal gyrus & TPOmid.R \\ 
  44 & R Calcarine fissure and surrounding cortex & CAL.R & 89 & L Inferior temporal gyrus & ITG.L \\ 
  45 & L Cuneus & CUN.L & 90 & R Inferior temporal gyrus & ITG.R \\ 
       \cline{4-6}
  46 & R Cuneus & CUN.R &  & \multicolumn{2}{l}{\textbf{Unclassified}}  \\ 
  47 & L Lingual gyrus & LING.L & 17 & L Rolandic operculum & ROL.L \\ 
  48 & R Lingual gyrus & LING.R & 18 & R Rolandic operculum & ROL.R \\ 
  49 & L Superior occipital lobe & SOG.L & 21 & L Olfactory cortex & OLF.L \\ 
  50 & R Superior occipital lobe & SOG.R & 22 & R Olfactory cortex & OLF.R \\ 
  51 & L Middle occipital lobe & MOG.L & 27 & L Gyrus rectus & REC.L \\ 
  52 & R Middle occipital lobe & MOG.R & 28 & R Gyrus rectus & REC.R \\ 
  53 & L Inferior occipital lobe & IOG.L & 41 & L Amygdala & AMYG.L \\ 
  54 & R Inferior occipital lobe & IOG.R & 42 & R Amygdala & AMYG.R \\ 
  55 & L Fusiform gyrus & FFG.L & 91 & L Cerebellum crus 1 & CRUS1.L \\ 
  56 & R Fusiform gyrus & FFG.R & 92 & R Cerebellum crus 1 & CRUS1.R \\ 
  \cline{1-3}
&  \multicolumn{2}{l}{\textbf{Attentional network}}  & 93 & L Cerebellum crus 2 & CRUS2.L \\ 
  3 & L Superior frontal gyrus, dorsolateral & SFGdor.L & 94 & R Cerebellum crus 2 & CRUS2.R \\ 
  4 & R Superior frontal gyrus, dorsolateral & SFGdor.R & 95 & L Cerebellum 3 & CB3.L \\ 
  5 & L Superior frontal gyrus, orbital & ORBsup.L & 96 & R Cerebellum 3 & CB3.R \\ 
  6 & R Superior frontal gyrus, orbital & ORBsup.R & 97 & L Cerebellum 4 5 & CB4\_5.L \\ 
  7 & L Middle frontal gyrus & MFG.L & 98 & R Cerebellum 4 5 & CB4\_5.R \\ 
  8 & R Middle frontal gyrus & MFG.R & 99 & L Cerebellum 6 & CB6.L \\ 
  9 & L Middle frontal gyrus, orbital & ORBmid.L & 100 & R Cerebellum 6 & CB6.R \\ 
  10 & R Middle frontal gyrus, orbital & ORBmid.R & 101 & L Cerebellum 7 & CB7b.L \\ 
  11 & L Inferior frontal gyrus, opercular & IFGoperc.L & 102 & R Cerebellum 7 & CB7b.R \\ 
  12 & R Inferior frontal gyrus, opercular & IFGoperc.R & 103 & L Cerebellum 8 & CB8.L \\ 
  13 & L Inferior frontal gyrus, triangular & IFGtriang.L & 104 & R Cerebellum 8 & CB8.R \\ 
  14 & R Inferior frontal gyrus, triangular & IFGtriang.R & 105 & L Cerebellum 9 & CB9.L \\ 
  15 & L Inferior frontal gyrus, orbital & ORBinf.L & 106 & R Cerebellum 9 & CB9.R \\ 
  16 & R Inferior frontal gyrus, orbital & ORBinf.R & 107 & L Cerebellum 10 & CB10.L \\ 
  29 & L Insula & INS.L & 108 & R Cerebellum 10 & CB10.R \\ 
  30 & R Insula & INS.R & 109 & Vermis 12 & VERMIS1\_2 \\ 
  59 & L Superior parietal gyrus & SPG.L & 110 & Vermis 3 & VERMIS3 \\ 
  60 & R Superior parietal gyrus & SPG.R & 111 & Vermis 4 5 & VERMIS4\_5 \\ 
  61 & L Inferior parietal gyrus & IPL.L & 112 & Vermis 6 & VERMIS6 \\ 
  62 & R Inferior parietal gyrus & IPL.R & 113 & Vermis 7 & VERMIS7 \\ 
  83 & L Temporal pole: superior temporal gyrus & TPOsup.L & 114 & Vermis 8 & VERMIS8 \\ 
  84 & R Temporal pole: superior temporal gyrus & TPOsup.R & 115 & Vermis 9 & VERMIS9 \\ 
  85 & L Middle temporal gyrus & MTG.L & 116 & Vermis 10 & VERMIS10 \\ 
  86 & R Middle temporal gyrus & MTG.R &  &  &  \\ 
  89 & L Inferior temporal gyrus & ITG.L &  &  &  \\ 
  90 & R Inferior temporal gyrus & ITG.R &  &  &  \\ 
   \hline \hline
\end{tabular}}
\caption{rs-fMRI data analysis. Regions of interest (ROIs) as defined by the AAL brain atlas and resting state networks (RSN).} 
\end{center}
\end{table}

\end{document}